\documentclass[12pt,preprint]{aastex}

\newcommand{\simlt}{\lower.5ex\hbox{$\; \buildrel < \over \sim \;$}}

\slugcomment{draft date: \today}

\begin{document}

\title{Molecular and Ionic shocks in the Supernova Remnant 3C~391}

\author{William T. Reach, Jeonghee Rho, and T. H. Jarrett}

\affil{Infrared Processing and Analysis Center, 
California Institute of Technology,
Pasadena, CA 91125}

\author{Pierre-Olivier Lagage}
\affil{Service d'Astrophysique, CEA, DSM, DAPNIA, Centre d'Etudes de Saclay,
F-91191 Gif-sur-Yvette cedex, France}

\email{reach@ipac.caltech.edu}

\def\textabstract{

New observations of the supernova remnant 3C391 are in the H2 2.12 micron and
[Fe II] 1.64 micron narrow-band filters at the Palomar 200-inch telescope, and
in the 5-15 micron CVF on ISOCAM. Shocked H2 emission was detected from the
region 3C391:BML, where broad millimeter CO and CS lines had previously been
detected. A new H2 clump was confirmed to have broad CO emission, demonstrating
that the near-infrared H2 images can trace previously undetected molecular
shocks. The [Fe II] emission has a significantly different distribution, being
brightest in the bright radio bar, at the interface between the  supernova
remnant and the giant molecular cloud, and following filaments in the radio
shell. The near-infrared [Fe II] and the mid-infrared 12-18 micron filter
images are the first images to reveal the radiative shell of 3C391. The
mid-infrared spectrum is dominated by bright ionic lines and H2 S(2) through
S(7). There are no aromatic hydrocarbons associated with the shocks, nor is
their any mid-infrared continuum, suggesting that macromolecules and very small
grains are destroyed. Comparing 3C391 to the better-studied IC443, both
remnants have molecular- and ionic-dominated regions; for 3C391, the
ionic-dominated region is the interface into the giant molecular cloud, showing
that the main bodies of giant molecular clouds contain significant regions 
with densities 100 to 1000/cm^3 and a small filling factor with higher-density.
The molecular shocked region  resolves into 16 clumps of H2 emission, with some
fainter diffuse emission but with no associated near-infrared continuum
sources. One of the clumps is coincident with a previously-detected OH 1720 MHz
maser. These clumps are interpreted as a cluster of pre-stellar, dense
molecular cores that are presently being shocked by the supernova blast wave. 

}

\begin{abstract}

New observations of the supernova remnant 3C~391 are presented in
the near-infrared, using the H$_2$ 2.12 $\mu$m and 
[\ion{Fe}{2}] 1.64 $\mu$m narrow-band filters in the Prime Focus
Infrared Camera on the Palomar Observatory Hale $200''$ telescope,
and in the mid-infrared, using the circular-variable filters in
the ISOCAM on the {\it Infrared Space Observatory}.
Shocked H$_2$ emission was detected from the region 3C~391:BML
($40''$ size),
where broad millimeter CO and CS lines had previously been detected.
A small H$_2$ clump, 45$''$ from the main body of 3C~391:BML, 
was confirmed to have broad CO emission, demonstrating that the near-infrared
H$_2$ images can trace previously undetected molecular shocks.
The [\ion{Fe}{2}] emission has a significantly different distribution,
being brightest in the bright radio bar, at the interface between the 
supernova remnant and the giant molecular cloud, and following
filaments in the radio shell. The near-infrared [\ion{Fe}{2}] and
the mid-infrared 12--18 $\mu$m filter (dominated by 
[\ion{Ne}{2}] and [\ion{Ne}{3}] images are the first images
to reveal the radiative shell of 3C~391.
The mid-infrared spectrum is dominated by bright ionic lines of 
[\ion{Fe}{2}] 5.5 $\mu$m, [\ion{Ar}{2}] 6.9 $\mu$m, 
[\ion{Ne}{2}] 12.8 $\mu$m, and [\ion{Ne}{3}] 15.5 $\mu$m,
as well as the series of pure rotational lines of H$_2$ S(2) 
through S(7).
There are no aromatic hydrocarbons associated with the
shocks, nor is their any mid-infrared continuum, suggesting
that macromolecules and very small grains are destroyed in the
shocks.
Comparing 3C~391 to the better-studied IC~443, both remnants have
molecular- and ionic-dominated regions; for 3C~391, the ionic-dominated
region is the interface into the giant molecular cloud, showing that
the main bodies of giant molecular clouds contain significant regions 
with densities $10^2$ to $10^3$ cm$^{-3}$ and a small filling factor of
higher-density regions.
The broad-molecular line region 3C~391:BML
was imaged in the 1-0 S(1) line at $1.5''$ resolution.
The molecular shocked region 
resolves into 16 clumps of H$_2$ emission, with some fainter diffuse
emission but with no associated near-infrared continuum sources. 
One of the clumps is coincident with a previously-detected OH 1720 MHz
maser to within our 0.3$''$ astrometry. These clumps are interpreted
as a cluster of pre-stellar, dense molecular cores that are presently being
shocked by the supernova blast wave. 

\keywords{supernova remnants, shock waves,
ISM: individual (3C~391)
}

\end{abstract}
                            

\section{Introduction}

The number of supernova remnants known to be interacting with
molecular clouds has increased 
over the last few years from one single
example, IC~443 \citep{denoyer443,white443}, to six:
IC~443,  W~51C \citep{koow51}, 3C~391 \citep{rr99}, 
W~44 \citep{setaw44} and W~28 \citep{arikawa},  HB~21 (G89.0+4.7)
\citep{koohb21}.  
The strong observational evidence for interactions includes broad CO lines,
bright far-infrared lines, or OH masers.
Other evidence can be infrared cooling lines from H$_2$, OH, and CO
tracing regions that are both hot and dense \citep{rr00,rho443}. 
The growing number of molecular interacting supernova remnants  
provides critical laboratories to study shock physics and comparison 
of the strong supernova shocks with other shocks in astrophysical 
objects such as wind shocks and outflows.
How the shocks develop around clouds and how shocks affect molecular cloud
evolution are not well understood. 
Chevalier (1999) considered two density regimes (cloud and intercloud) 
for the pre-shock gas. The observed infrared lines require,
in addition to the intercloud medium, shocks into 
gas with both moderate ($\sim$10$^{2}$ cm$^{-3}$) and high ($\sim$ 
10$^{4}$ cm$^{-3}$) pre-shock densities. 
The results suggest that an ensemble of different 
shocks develops on small scales, due to the high contrasts in density.

In the case of IC~443, studies have shown that there are two types of
shocks, but at any given place within the remnant, only a single type of
shock is evident \citep{rho443}. 
The {\it 2MASS} image of IC~443 shows a striking contrast in 
near-infrared color between the south (where
the shock is impacting dense molecular gas) and
the northeast (where the shock is impacting lower-density material).
The northeast rim is mostly [\ion{Fe}{2}] line 
emission in the near-infrared, with strong [\ion{O}{1}] and other ionic lines
in the mid- and far-infrared,
consistent with a J-shock with a shock velocity of 100 km s$^{-1}$ and a 
pre-shock density of order $10^2$ cm$^{-3}$.
In contrast, the southern ridge
is dominated by H$_2$ line emission exhibiting 
a clumped and  knotty structure. A surprising result is that no
atomic lines were not detected from the southern ridge 
in recent ISOCAM observations \citep{cesarsky443}.
The ISOCAM observations are consistent with a pure C-shock with 
a low shock velocity of 30 km s$^{-1}$ and a dense medium (10$^{4}$ cm$^{-3}$).
Because IC~443 is a uniquely well-studied molecular-interacting remnant, 
it is not clear whether the interstellar medium conditions were special 
around the IC~443 progenitor (recently located by \citet{olbert}), 
or whether the strong variations in 
pre-shock conditions are characteristic of the environments of Type 2
supernovae.

\begin{figure}
\plotone{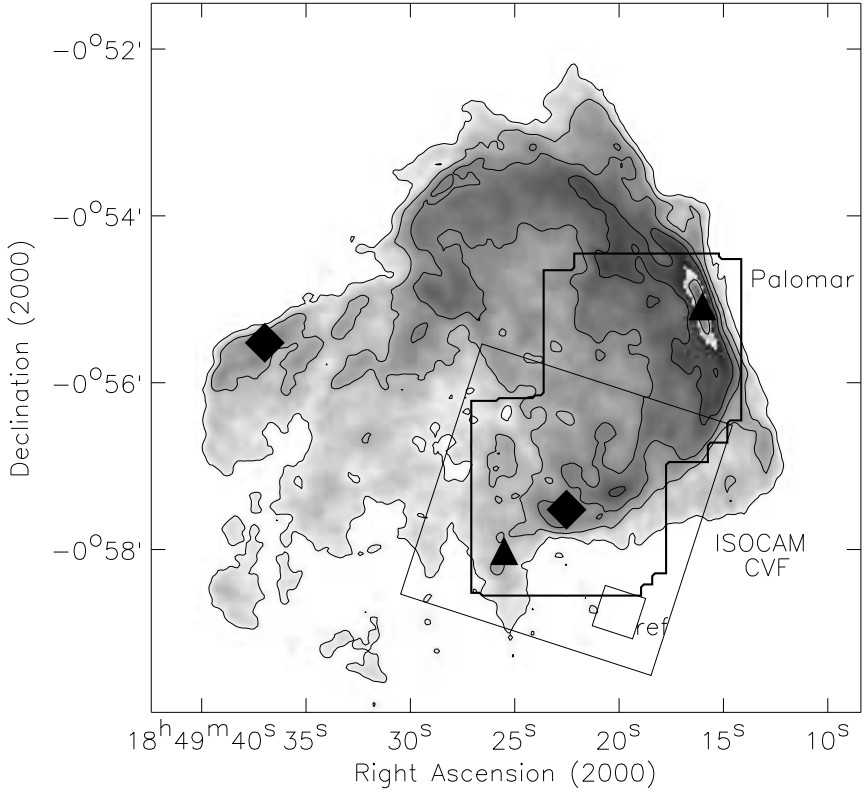}
\caption{A finder chart for the regions observed in 3C~391. The greyscale
and thin contours are the 1.4 GHz radio continuum image
\citep{reymof}, showing the bright 
radio bar in the northwest, where the shock front is striking the surface
of a molecular cloud, and fainter emission extending to the southeast,
where the remnant is expanding into a lower-density medium. 
The greyscale runs from white (faint) to 
black (medium), then back to white and up to black again in
the radio bar (very bright).
The large, black square with $18^\circ$
rotation outlines the region observed with the ISOCAM CVF.
The small, black square just inside the southwest corner of the ISOCAM field 
indicates the reference position, where the line emission was 
weakest in the ISOCAM data.
The thick outline shows the region observed
with the Palomar Observatory $200''$ telescope; this region is
the combination of two, dithered pointings.
The two filled diamonds mark the locations of 
the OH 1720 MHz masers. 
The two filled triangles mark the location of the peak [\protect\ion{Fe}{2}]
emission (northwest, used in Fig.~\ref{cormass}a) and the position 
of an ionic shock within the ISOCAM region (south, used in Fig.~\ref{ionspec}).
\label{finder}}
\end{figure}

In this paper, we look in some detail at the morphology of molecular and ionic
shocks in the supernova remnant 3C~391 (G31.9+0.0).
Figure~\ref{finder} shows the radio image of 3C~391. 
The remnant has a distinctive 
`breakout' radio morphology, suggesting that it has struck a region of
significantly higher density northwest of the explosion site \citep{reymof}.
The X-ray emission is contained within the radio shell, making 3C~391 one
of the new class of `mixed-morphology' remnants with bright radio shells
surrounding diffuse, central, thermal X-rays \citep{rho391,RP98}.
Millimeter-wave CO observations revealed that there is indeed a giant molecular
cloud to the northwest of the remnant and at a rotation-curve distance
consistent with the estimated distance of the remnant \citep{wilner}.
Two OH 1720 MHz masers were detected from 3C~391, suggesting that in at least
two places the shock may be interacting with molecular gas \citep{Frail96}.
The remnant is so bright in the [\ion{O}{1}] line that it appears that relatively
dense shocks are present both toward the OH masers and the bright radio bar
\citep{rr96}. One of the OH masers was found to be included in a region of broad 
CS and CO lines, clearly revealing the shocked gas, but somewhat mysteriously only
revealing it from a small region in the southern part of the remnant;
this `broad molecular line' region is called 3C~391:BML \citep{rr99}.

The nature of 3C~391:BML is unknown because it
has not been imaged with sufficient resolution. 
Based on the millimeter-wave observations ($13''$ beam), a high cloud pressure 
was inferred,  suggesting gravitational compression is needed to
keep the cloud from rapidly dissipating; the self-gravity also suggests 
imminent collapse \citep{rr99}.
The detection of far-infrared emission from H$_2$O, CO, and OH, in an $80''$ beam, 
further suggested the presence of high-density gas.
But neither the structure of the cloud (highly fragmented or monolithic) nor
the chemistry (diffuse gas with low H$_2$O abundance, or smaller clumps
with the predicted high H$_2$O post-shock abundance) could be resolved.
The new observations presented in this paper explore the distribution of shocked
molecular and ionized gas, using spatially-resolved spectroscopy and narrow-band
imaging, revealing dramatic color differences between the different parts of
the remnant, and uncovering a cluster of pre-stellar molecular cores in 3C~391:BML.

\section{Observations}

The primary observations for this paper are two sets of images, one
made with the {\it Infrared Space Observatory} and the other with 
Palomar 200$''$ telescope. 
Figure~\ref{finder} illustrates the regions we observed, in the context
of the entire supernova remnant, as traced by the
radio image \citep{reymof}. There are also two `geographical'
landmarks in the region covered, to which we will often refer.
First is the location of the radio-bright bar, at the northwestern  
edge of the remnant where the shock is impinging on the front of a 
molecular cloud \citep{wilner}; the location of the radio peak is clear in
Fig.~\ref{finder}. Second is the location of
the southern OH maser \citep{Frail96},
indicated by a $\star$ in Fig.~\ref{finder},
where the shock is impinging
on a dense clump \citep{rr99}.

\subsection{{\it Infrared Space Observatory}}

\begin{figure}
\plotone{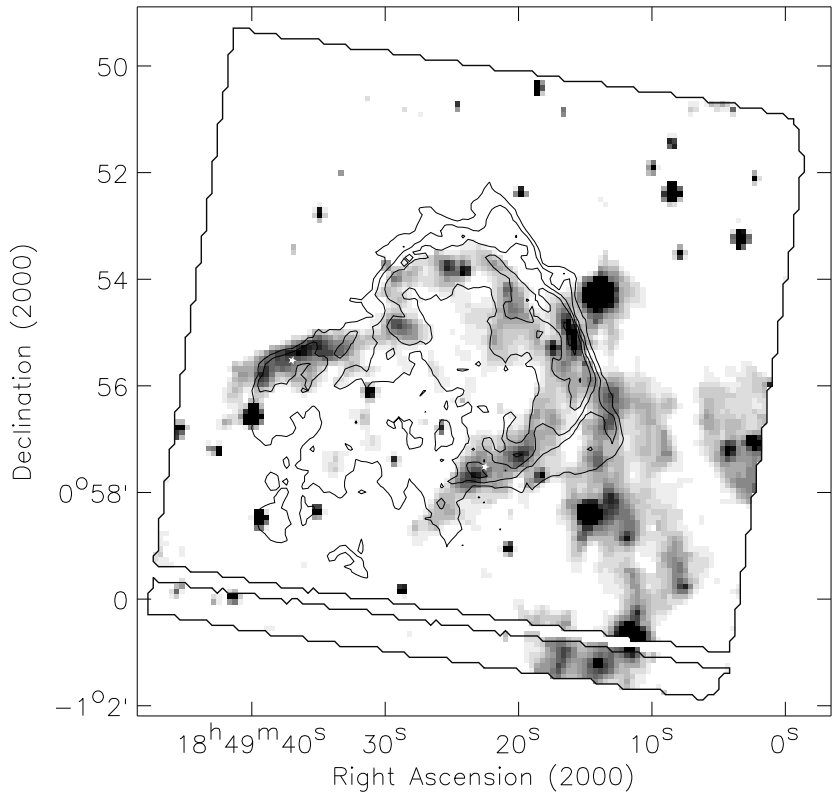}
\caption{ISOCAM 12--18 $\mu$m filter image of 3C~391. The radio
contours and symbols (same as Fig. 1) are overlaid for comparison. 
This the first infrared image of the 3C~391, which is clearly visible
as the diffuse, filamentary emission roughly following the radio shell.
\label{figlw3}}
\end{figure}

Observations of 3C~391 were made using the {\it Infrared Space Observatory}
\citep{kessler} mid-infrared camera ISOCAM \citep{ccesarsky} through a
variety of filters, in order to determine the morphology and physical origin
of infrared emission from 3C~391. ISOCAM is a $32\times 32$ pixel array,
and we selected the 6$''$ pixel field of view to get a large, un-vignetted
field of view. First, on April 11, 1997, an image of the entire remnant was
made in the wide LW3 (12--18 $\mu$m) filter. The mosaic was made with
a $7\times 7$ raster of pointings separated by 80$''$, which is much
smaller than the 192$''$ instantaneous field of view to provide
redundancy. At each pointing, a set of 7 frames of 2.1 sec duration
were taken, providing another level of redundancy. The observation
lasted 36 minutes. Cosmic rays
were identified (and masked) using a temporal high-pass filter,
and the pixel-to-pixel gain was determined from the un-registered
median of all of the images; both steps make use of the high level
of redundancy in the data. Figure~\ref{figlw3} shows the 12--18 $\mu$m image,
which covers the entire remnant and is dominated by remnant emission
and about 100 stars. 
The {\it IRAS} data for 3C~391 are hopelessly contaminated by stars,
such as the very bright star just outside the radio bar, 
which swamp the remnant at low angular and spectral resolution.
The ISOCAM image shows that mid-infrared emission from 3C~391
generally follows the radio shell. The nature of the 12--18 $\mu$m emission
from the remnant is discussed below in \S 5. 

There is an extension of diffuse 12--18 $\mu$m emission,
extending southward from the radio bar, toward lower right 
of Figure~\ref{figlw3}. 
It is not clear whether this extension is related to the remnant, although
it appears, from the 12--18 $\mu$m image alone, to be connected to it.
The extension does not have associated radio emission, so it is well outside
what we thought of as the boundary of the remnant.
Instead, the extension may be a
photodissociation region on the surface of the same molecular cloud with which 
the remnant is interacting; in this case the emission would be due to 
aromatic hydrocarbons and fine structure lines of [\ion{Ne}{2}] and [\ion{Ne}{3}].

Having demonstrated that 3C~391 is a bright mid-infrared source,
we returned to 3C~391 on November 3, 1997, to obtain a complete
set of images with the circular-variable filter (CVF). The
goal was to discern what is the emission mechanism: lines or continuum.
For these CVF
observations, only a single field of view (192$''$) was observed,
centered on the location of the bright OH maser and wide CO lines.
The observation proceeded by setting the filter wheel near the highest
wavelength of the long-wavelength CVF [16.14 $\mu$m], taking 12 frames of 2.1 sec duration,
moving down to the next step of the long-wavelength CVF [16.05 $\mu$m], etc.,
until near the end of the long-wavelength CVF [10.09 $\mu$m; 69 steps]
then setting the filter wheel near the highest wavelength of the
next CVF [9.986 $\mu$m] and until near the end of this CVF
[5.14 $\mu$m; 81 steps]. The entire CVF observation took 68 minutes
to complete. The data were reduced using the same methods
described in another recent paper \citep{smcpah}.
A correction was applied to remove the transient gain variations,
based on the IAS method \citep{transient}, which led to 
typically $\sim 10$\% changes in the surface brightness, 
with $\sim 20$\% changes at wavelengths where there are sharp changes
in the surface brightness. Before we applied the transient gain
correction, there were some differences between the spectra
taking in the increasing and decreasing wavelength directions,
due to subtle differences in
the time constants of the detectors; these remain problematic only
around the 6.2 and 11.3 $\mu$m PAH features. 
The final reduction yielded
a cube of $32\times 32$ spatial pixels (with one column dead)
$\times 147$ unique wavelengths. The spectral passband of each CVF
step is 0.33 $\mu$m wide at 16 $\mu$m, 0.27 $\mu$m wide
at 10 $\mu$m; 0.21 $\mu$m wide at 9 $\mu$m, and 0.14 $\mu$m wide at
5.2 $\mu$m. 

\begin{table}
\caption[]{Line brightnesses$^a$ for 3C~391:BML}\label{cvfline} 
\begin{flushleft} 
\begin{tabular}{lcccccc} 
\hline
species & transition & $\lambda$  & $10^4 I$ & $10^{4} I^\prime$ \\
        &            &  ($\mu$m)  &
	\multicolumn{2}{c}{(erg~s$^{-1}$~cm$^{-2}$sr$^{-1}$)} \\
\hline
ISOCAM$^b$ & & & & \\
\hline
Ne$^{++}$       &    &  15.56   &2.11 &  2.56\\
Ne$^+$  &            &  12.81   &2.89 &  3.95\\
H$_2$   & 0-0 S(2)   &  12.28   &1.07 &  1.46\\
H$_2$   & 0-0 S(3)   &   9.66   &1.89 &  6.39\\
H$_2$   & 0-0 S(4)   &   8.03   &2.45 &  3.66\\
Ar$^+$   &           &   6.99   &4.87 &  6.00\\
H$_2$   & 0-0 S(5)   &   6.91   &9.84 & 12.13\\
H$_2$   & 0-0 S(6)   &   6.11   &3.82 &  4.79\\
H$_2$   & 0-0 S(7)   &   5.51   &6.21 &  8.06\\
Fe$^+$   &           &   5.34   &3.91 &  5.08\\
\hline
Palomar$^c$ & & & & \\
\hline
H$_2$   & 1-0 S(1)   &   2.12   & 2.1   & 14.5 & \\
Fe$^+$   &           &   1.64   & 3.0   & $\sim$40$^d$ \\
\hline
\end{tabular} 
\noindent\par $^a$ Surface brightness observed is $I$, and corrected for 
extinction ($A_V=19$) is $I^\prime$
\noindent\par $^b$ Brightness in the $6''$ peak pixel of the CVF image of 3C~391:BML
\noindent\par $^c$ For H$_2$, brightness of typical clumps in 3C~391:BML;
for Fe$^+$, brightness averaged over the $10''\times 26''$ radio bar.
\noindent\par $^d$ This value is uncertain because the extinction correction for 
the near-infrared [\ion{Fe}{2}] line is large.
\end{flushleft} 
\end{table}  

\begin{figure}
\epsscale{1}
\plotone{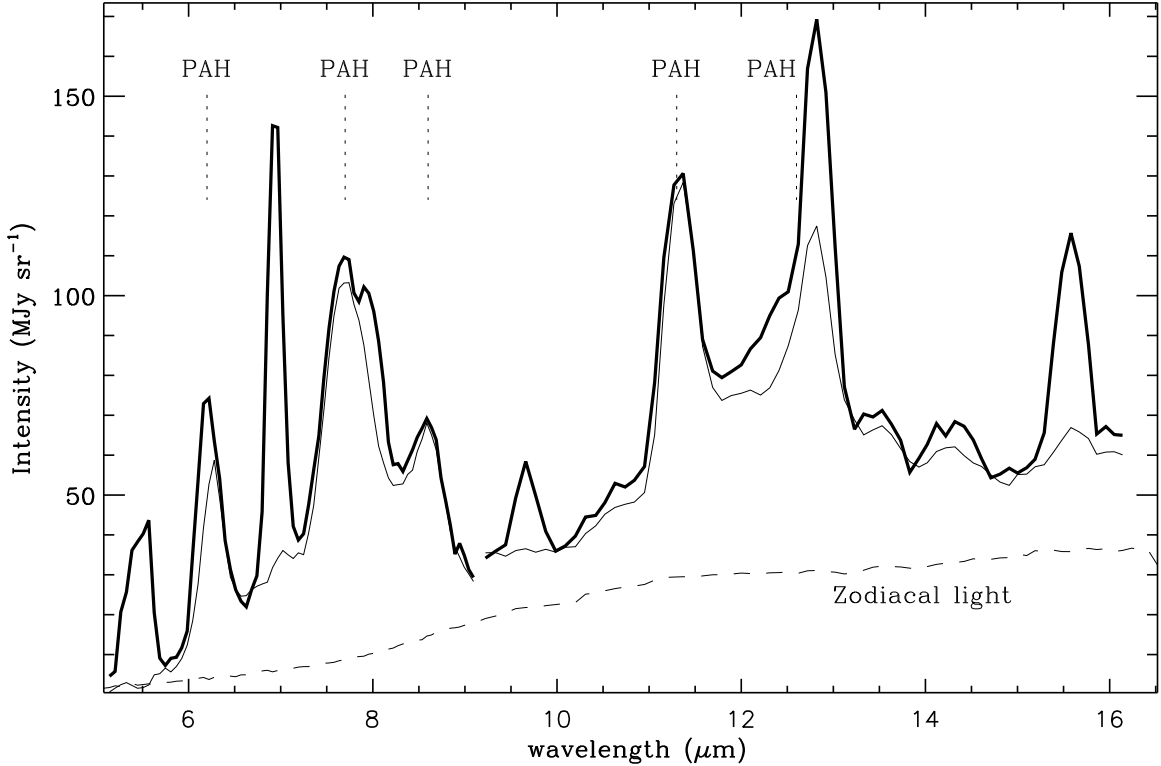}
\caption{The mid-infrared spectrum of a single 6$''$ ISOCAM CVF pixel
(18$^h$ 49$^m$ 23.3$^s$  -00$^\circ$ 57$'$ 34$''$ [J2000],
southern filled diamond in Fig.~\ref{finder}.)
The thick line shows the total surface brightness versus
wavelength for this pixel. 
Most of the observed surface brightness is from a combination of
zodiacal light and unrelated Galactic emission.
The thin dashed line shows the 
predicted, absolute spectrum of the zodiacal light for this observation.
The thin solid line shows the spectrum averaged over an area in the 
CVF observation with the least [\protect\ion{Ne}{3}] and radio continuum emission.
(The reference area is centered on 18$^h$49$^m$20$^s$ -0$^\circ$58$'$45$''$
and outlined in the southwest corner of Fig.~\ref{finder}.)
The broad features (labeled `PAH') 
at 6.2, 7.7, 8.6, 11.3, and 12.6 $\mu$m are present
throughout the observed region, with nearly constant brightness
(hence the good match between thick and thin curves);
these features appear to be unrelated to the supernova remnant.
The wavelengths where the thick curve deviates significantly from the thin
curve correspond to ground-state ionic fine structure lines and
and H$_2$ pure rotational lines.
\label{cvfspec}}
\end{figure}

\begin{figure}
\epsscale{1}
\plotone{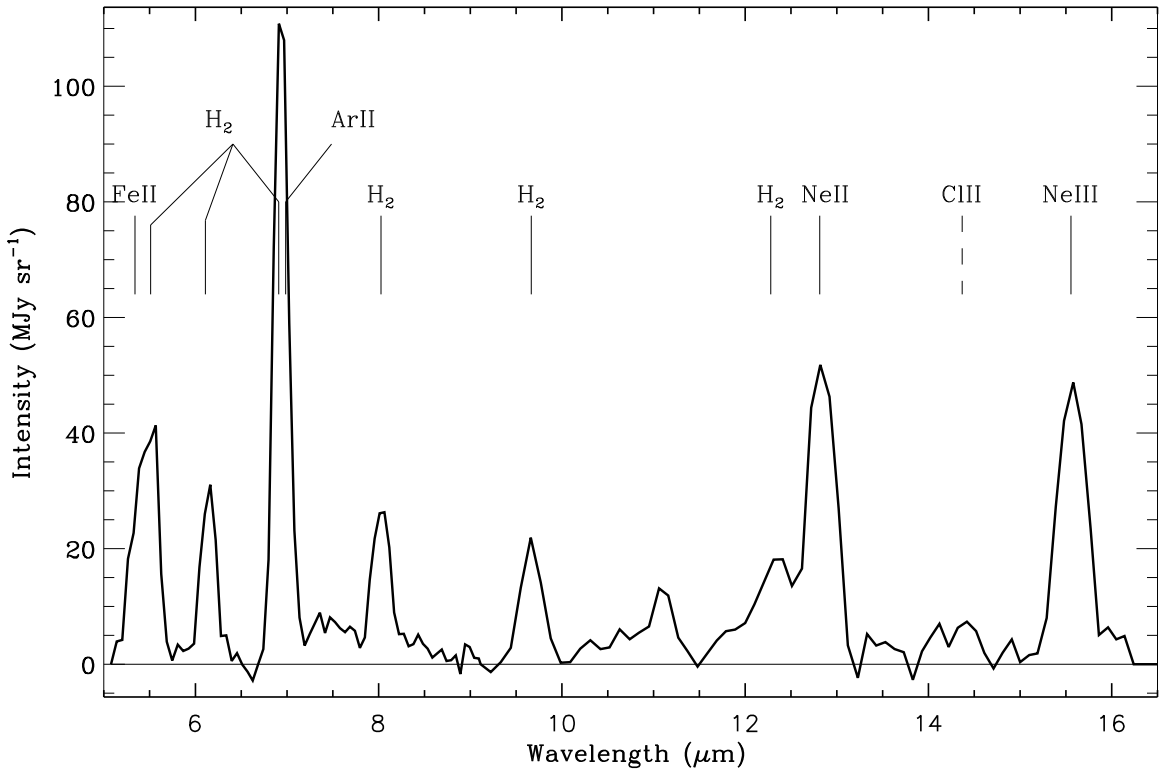}
\caption{ISOCAM CVF spectrum of the molecular shock front 3C~391:BML, obtained
by subtracting the `on' and reference spectra
(thick and thin solid curves in Fig.~\ref{cvfspec}).
This difference spectrum is completely dominated by narrow emission lines:
the continuum (and broad PAH features) are unrelated to the remnant.
Each emission line is labeled with the ion or molecule that produced it.
The `wiggle' around 11.3 $\mu$m is an artifact of subtraction of the
reference spectrum at the location of the very bright 11.3 $\mu$m
aromatic hydrocarbon line, due to slight changes in temporal response for
the detectors making the `on' and reference spectra.
\label{clumpspec}}
\end{figure}

The spectrum of the shocked gas was separated from that of the foreground emission
as follows.
Figure~\ref{cvfspec} shows the spectra of the molecular peak and a reference
position.
Figure~\ref{clumpspec} shows the difference between the spectrum toward
the peak of the line emission and the reference spectrum.
The reference spectrum removes unrelated, foreground and background,
emission from the remnant spectra.
Most of the observed surface brightness is
effectively removed by taking the difference, because most of the
surface brightness is due to smooth, foreground emission from
the zodiacal light (which is extremely smooth) and the diffuse
galactic emission.
The difference spectrum is completely dominated by narrow spectral features,
including 5 bright lines of H$_2$ and 4 bright atomic fine-structure lines.
All of the features in this spectrum, brighter than 
10 MJy~sr$^{-1}$ can be identified with known atomic and molecular
spectral lines, except for the `wiggle' near 11.3 $\mu$m which is a
residual subtraction error from the bright galactic feature at that
wavelength.

Table~\ref{cvfline} lists the line brightnesses from the ISOCAM data.
The line brightnesses were measured from the same spectrum shown in
Figure~\ref{clumpspec}, using the spectral response profiles of
the circular variable filter. The extinction toward 3C~391 is very 
high, owing to its location in the galactic plane and its great distance.
Spectral analysis of the X-ray data yielded a foreground column
density of 2--$3.6\times 10^{22}$~cm$^{-2}$, with the higher numbers
applying in the northwestern region \citep{rho391}. The X-ray-derived
column densities are probably on the low size, as they are averages over
relatively large regions. Thus, we will use a nominal foreground
column density of $3.6\times 10^{22}$~cm$^{-2}$, corresponding to a
visual extinction of $A_V=19$ mag. Each line brightness was then
corrected for extinction \citep{riekelebofsky}, and the corrected
values are listed in the last column of Table~\ref{cvfline}.

\begin{figure}
\epsscale{1}
\plotone{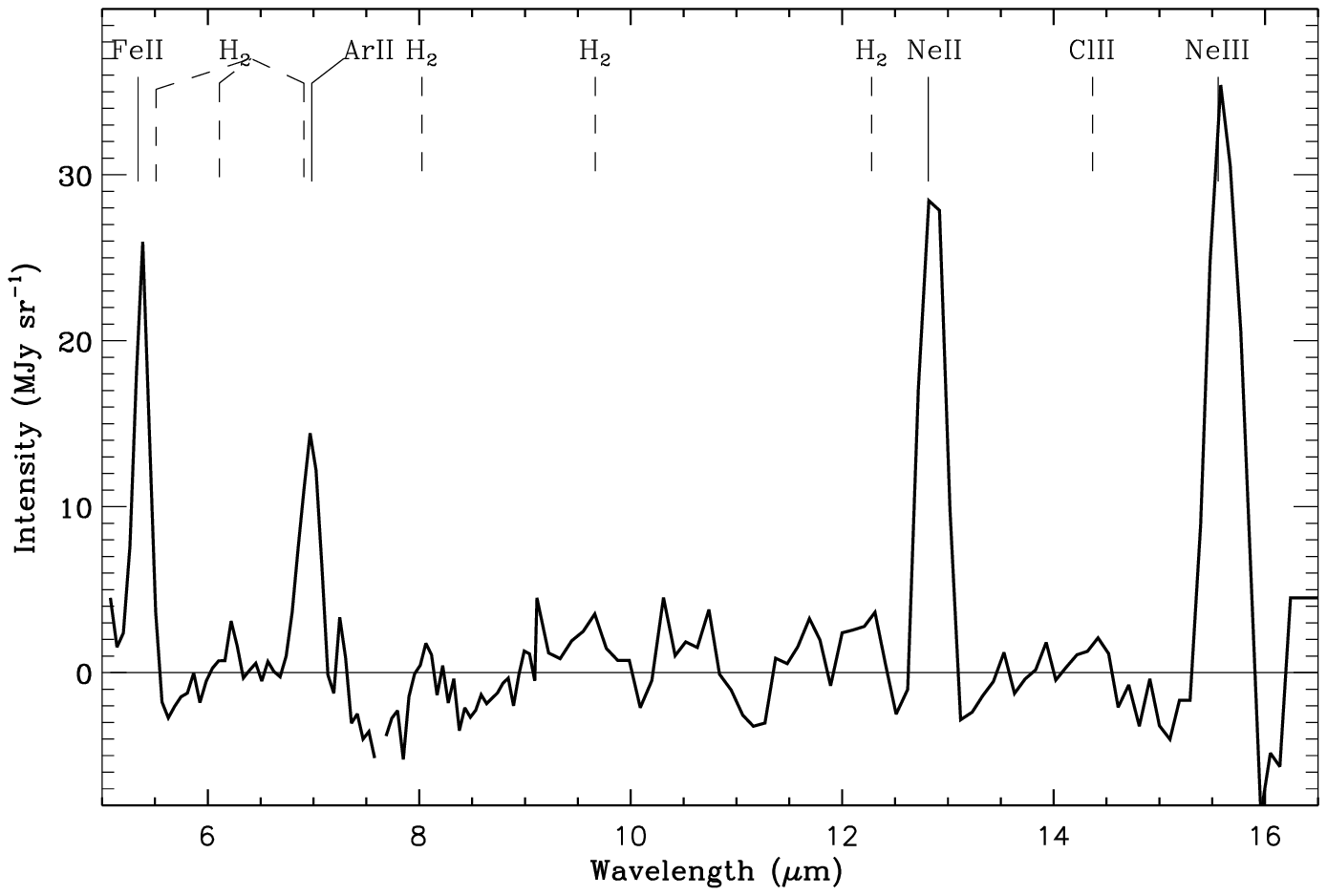}
\caption{ISOCAM CVF spectrum of an ionic shock position near 3C~391:BML, 
obtained by subtracting the reference spectrum from the spectrum of a single pixel on 
the radio and ionic shell but away from the molecular emission.
(The position is shown as the southern, filled triangle in Fig.~\ref{finder}.)
This spectrum is completely dominated by narrow emission lines from atomic ions:
neither continuum nor aromatic features are present.
\label{ionspec}}
\end{figure}

The mid-infrared spectrum of an ionic shock in 3C~391 can be separated form
that of the molecular shock, because the molecular shock is spatially
confined only to the 3C~391:BML region. Figure~\ref{ionspec} shows the
ISOCAM CVF spectrum, after subtracting the reference spectrum outside the
remnant, of a single pixel, at 18$^h$49$^m$25.5$^s$, -00$^\circ$ 58$'$ 01$''$ 
(see Fig.~\ref{finder}) that is on the radio shell but away from
the molecular region. Comparing this
spectrum to Fig.~\ref{clumpspec}, we see that the ionic shock is
completely different from the molecular shock, which is only $54''$ away.
The very different spectral shapes confirm our identification of
the blended [\ion{Fe}{2}] and H$_2$ lines near 5.5 $\mu$m as well as the
blended [\ion{Ar}{2}] and H$_2$ lines near 6.9 $\mu$m. In both cases, the
line shifts (and dims) from the blended lines (in Fig.~\ref{clumpspec})
to the ionic contribution only (in Fig.~\ref{ionspec}).

\subsection{Palomar Observatory}

\begin{figure}
\epsscale{1}
\plotone{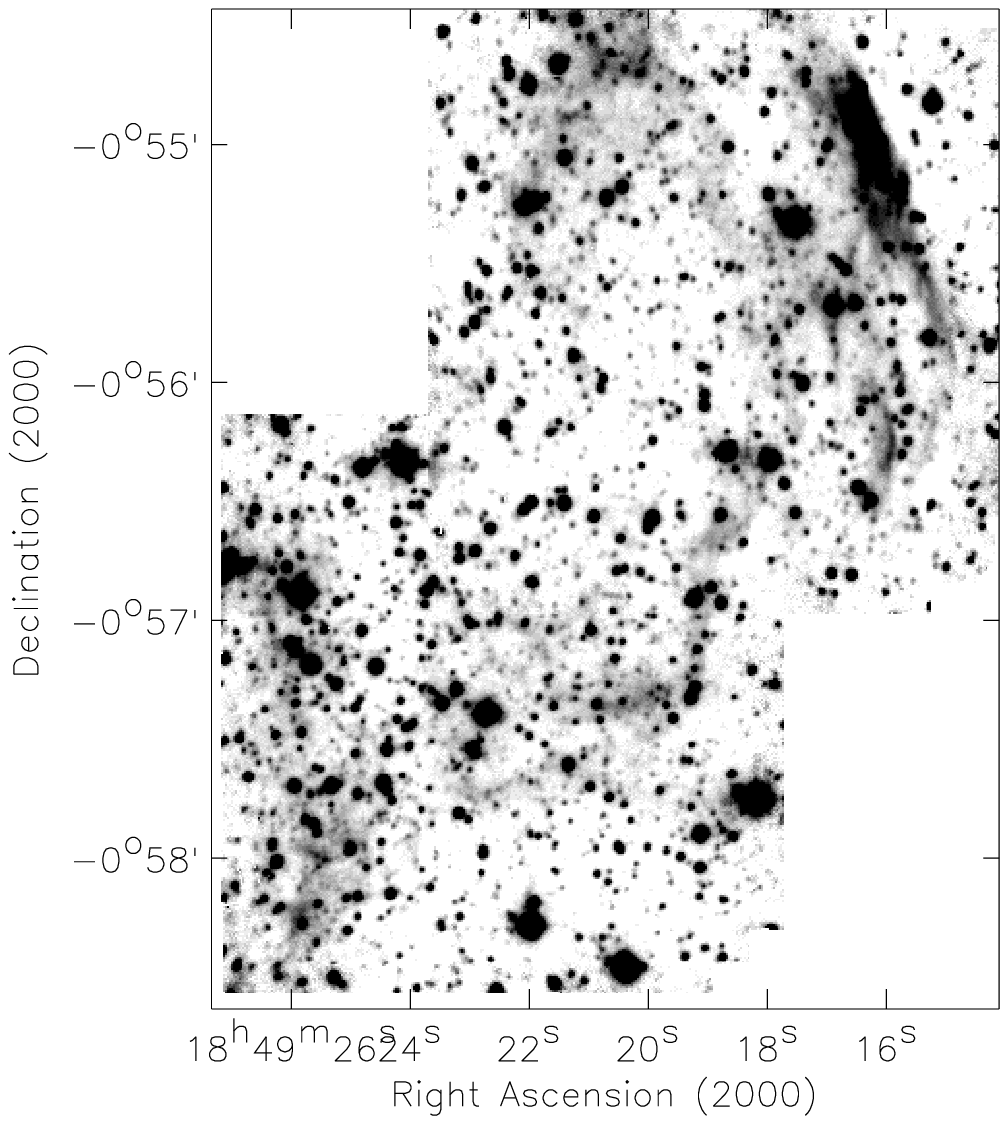}
\caption{PFIRCAM image of the [\protect\ion{Fe}{2}] 1.644 $\mu$m emission
from 3C~391. The portion of the remnant covered by this image is shown in
Fig. 1. The bright [\protect\ion{Fe}{2}] emission is located precisely within the
bright radio bar in the northwestern part of the remnant. Fainter
[\protect\ion{Fe}{2}] emission is spread over the remnant, generally outlining
the radio shell. The faint stars detected in this image are
approximately 45 $\mu$Jy (18.4 mag), and the faint diffuse emission
in this image is approximately $2.5\times 10^{-5}$ erg~s$^{-1}$~cm$^{-2}$~sr$^{-1}$.
The seeing in this image is approximately $1.2''$ (FWHM).
\label{fepal}}
\end{figure}

\begin{figure}
\epsscale{1}
\plotone{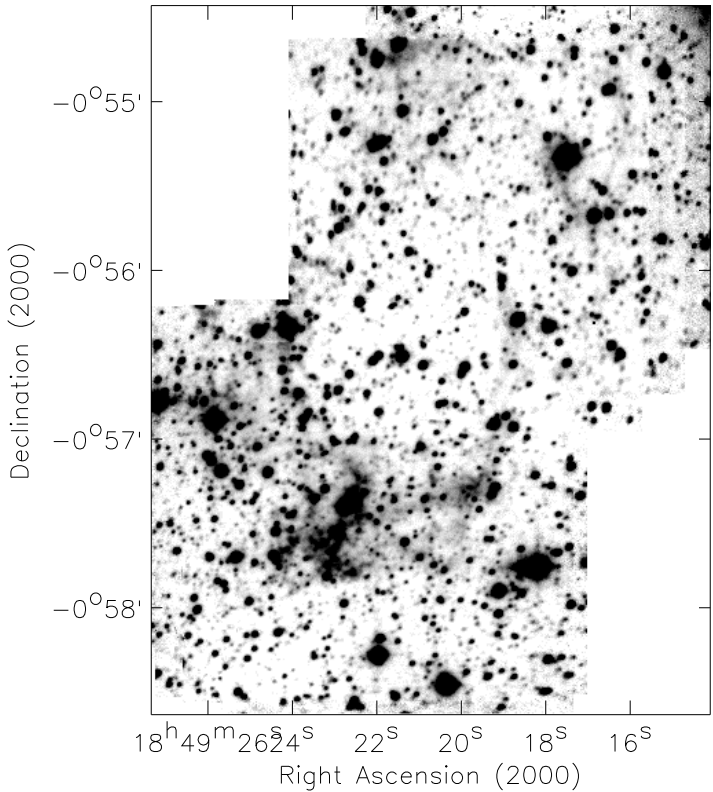}
\caption{PFIRCAM image of the H$_2$ $v=1-0$ S(1) 2.12 $\mu$m emission
from 3C~391. The portion of the remnant covered by this image is shown in
Fig. 1. The H$_2$ emission is mostly confined to a small, $\sim 20''$ region
in the southern part of the remnant, near the position of the broad
radio molecular lines and OH maser. Some very faint emission is extended over
the remnant, but it is near the limit of sensitivity of this image.
The faint stars detected in this image are
approximately 70 $\mu$Jy (17.5 mag), and the faint diffuse emission
in this image is approximately $7\times 10^{-5}$ erg~s$^{-1}$~cm$^{-2}$~sr$^{-1}$.
The seeing in this image is approximately $1.5''$ (FWHM).
\label{h2pal}}
\end{figure}

On July 16-17, 2000, we observed 3C~391 using the 
Prime Focus Infrared Camera (PFIRCAM)
on the Hale 200-inch telescope on Mount Palomar. 
The PFIRCAM has a $256\times 256$ pixel
array, with a pixel scale of 0.494$''$ at the f/3.3 prime focus of 
the 200-inch telescope. 
The seeing was fair during these observations, ranging 
from 1.3$''$ to 2$''$.
In order to remove temporally varying sky emission, we beam-switched between
the remnant and a nearby reference position, shifting by small dithers for
each on-off pair. 
The `reference' position was chosen by inspecting the 2MASS images and
finding a local minimum in the density of bright stars; it was then adjusted
at the telescope to further avoid bright stars.
The dithered `reference' observations were combined to generate a median 
sky image, which was subtracted from the `on' observations. The sky-subtracted
`on' observations were then combined into the final mosaic images.
Figures~\ref{fepal} and ~\ref{h2pal} show the PFIRCAM 
images of 3C~391 in two filters: [\ion{Fe}{2}] 1.644 $\mu$m and
H$_2$ $v=1\rightarrow 0$ S(1) 2.12 $\mu$m. 

The astrometry and absolute calibration of the PFIRCAM 200-inch images 
were determined by comparison with the Point Source Catalog of the 
2 Micron All-Sky Survey, 2MASS \citep{twomassref}. 
The 2MASS astrometry is accurate to 0.2$''$, and the photometry is
accurate to 5\%. The rms deviations between
the astrometric solutions and the 2MASS catalog for 12 stars
are 0.08$''$ for the H$_2$ image and 0.12$''$ for the [\ion{Fe}{2}] image.
The absolute calibration of the images was determined as
follows. First, the flux density per data number (DN)
was determined for stars fainter than 0.2 Jy using the average
flux density (from the 2MASS catalog) per integrated DN in the
PFIRCAM image. 
\def\extra{
For brighter stars,
the nonlinearity was adequately fit by a simple model
\begin{equation}
F_\nu = \left\{ \begin{array}{ll}
     k D & \mbox{for $D \le D_0$} \\
     k D \left[1 + \frac{(D-D_0)^2}{C D}\right] &
        \mbox{for $D > D_0$}
	\end{array}
     \right.
\end{equation}
where the parameter $k$ is determined from faint stars and $C$ is
determined from bright stars.
}
The surface brightness of a single pixel is then determined from the 
raw image DN using $I=k D \Delta\nu/\Omega$, where
$\Omega$ is the pixel solid angle.
Table~\ref{pfircal} gives the filter bandwidths, $\Delta\nu$.
The calibration factor, $k$, was determined using a least-squares
fit between the Palomar and 2MASS photometry; for the H$_2$ image,
we used 61 stars with K$_s$ magnitudes 9.5--12, and for the [\ion{Fe}{2}]
image, we used 58 stars with H magnitudes 10.2--13.1.
Table~\ref{pfircal} summarizes the calibration parameters.
Our absolute calibration scheme is different from the common usage
of `standard stars' observed occasionally during the night and compared to
standard references. Instead, the 2MASS data form a set of standards
distributed widely across the sky and absolutely calibrated
\footnote{For details, see the 2MASS Explanatory Supplement, Cutri et al.,
currently available at 
{\small\tt http://www.ipac.caltech.edu/2mass/releases/second/doc/explsup.html}}.
The obvious advantage of our scheme is that the
calibration data (stars) and the target data (diffuse emission) were taken
in the exact same manner along the same path through the atmosphere. 
The weather was not completely clear (high-altitude haze for part of the night) 
while taking the H$_2$ data, but our absolute calibration
will still be accurate. All near-infrared observations of fields
containing 2MASS sources are therefore `photometric' and can
be absolutely calibrated.


\begin{table}
\caption[]{Calibration of PFIRCAM filters}\label{pfircal} 
\begin{flushleft} 
\begin{tabular}{lcccc} 
\hline
filter & $\lambda$  & $\Delta\nu/\nu$ & $k$     & $k\Delta\nu/\Omega$  \\
       &   ($\mu$m) &              &($\mu$Jy/DN)&(erg~s$^{-1}$~cm$^{-2}$~sr$^{-1}$~DN$^{-1}$) \\
\hline
H$_2$  & 2.12       & 1\%            & 2.05     & $5.05\times 10^{-6}$  \\
\mbox{[\protect\ion{Fe}{2}]} & 1.644 & 1\%   & 1.59     & $5.04\times 10^{-6}$  \\
H$_2$ cont & 2.19   & 1\%            & 1.18     & ... \\
\hline
\end{tabular} 
\end{flushleft} 
\end{table}  

\begin{figure}
\epsscale{1}
\plotone{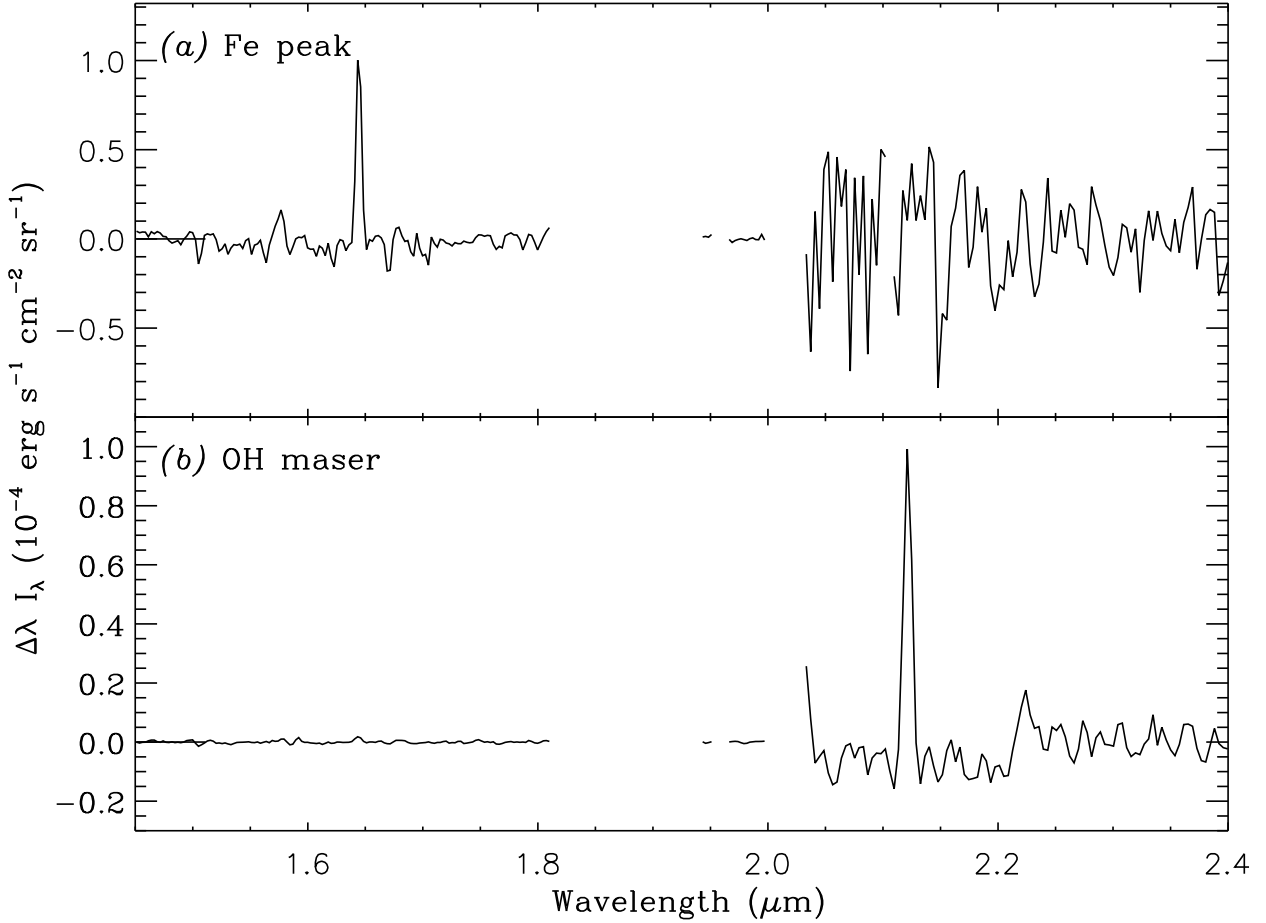}
\caption{Palomar 60$''$ H and K$_s$-band spectra toward two positions in 3C~391. 
{\it (a)} The upper panel shows the spectrum toward the northwest, where the radio 
emission peaks, at the interface
between the remnant and the molecular cloud. The [\protect\ion{Fe}{2}] line at 1.644 $\mu$m
is very bright, and no other lines are detected. 
(The position is indicated by the northwestern filled triangle in 
Fig.~\ref{finder}.)
{\it (b)} The lower panel shows the spectrum toward the OH maser, which is part
of the H$_2$ mission complex in the southern part of the remnant. The H$_2$ $v=1-0$
S(1) (2.1218 $\mu$m)and S(0) (2.2235 $\mu$m) lines dominates in the K$_s$ band, with a 
weak [\protect\ion{Fe}{2}] 1.644 $\mu$m
line in the H band. Neither spectrum shows a continuum, and neither spectrum
shows evidence for Br$\gamma$ (2.166 $\mu$m).
\label{cormass}}
\end{figure}

While writing this paper, we obtained two near-infrared spectra of 3C~391
using the CORMASS instrument \citep{wilson} on the Palomar 60$''$ telescope.
These spectra were calibrated by comparing the observed brightness of
the detected spectral lines to the narrow-band images from the 200$''$ of
the same location. These spectra clearly illustrate which are
the dominant emission lines in the near-infrared. 
Figure~\ref{cormass} shows the spectra toward the radio peak (top panel)
and the OH maser (bottom panel). 
Comparing these two spectra clearly demonstrates
that the location of bright [\ion{Fe}{2}] in Fig.~\ref{fepal} is
completely dominated by Fe, while the location of bright H$_2$
in Fig.~\ref{h2pal} is completely dominated by H$_2$, with negligible
continuum in the narrow-band filters. Further,
the two spectral lines we observed with PFIRCAM are, by far, the
brightest lines in the H and K near-infrared bands.

\section{Comparison of Molecular and Ionic images}

\begin{figure}
\plotone{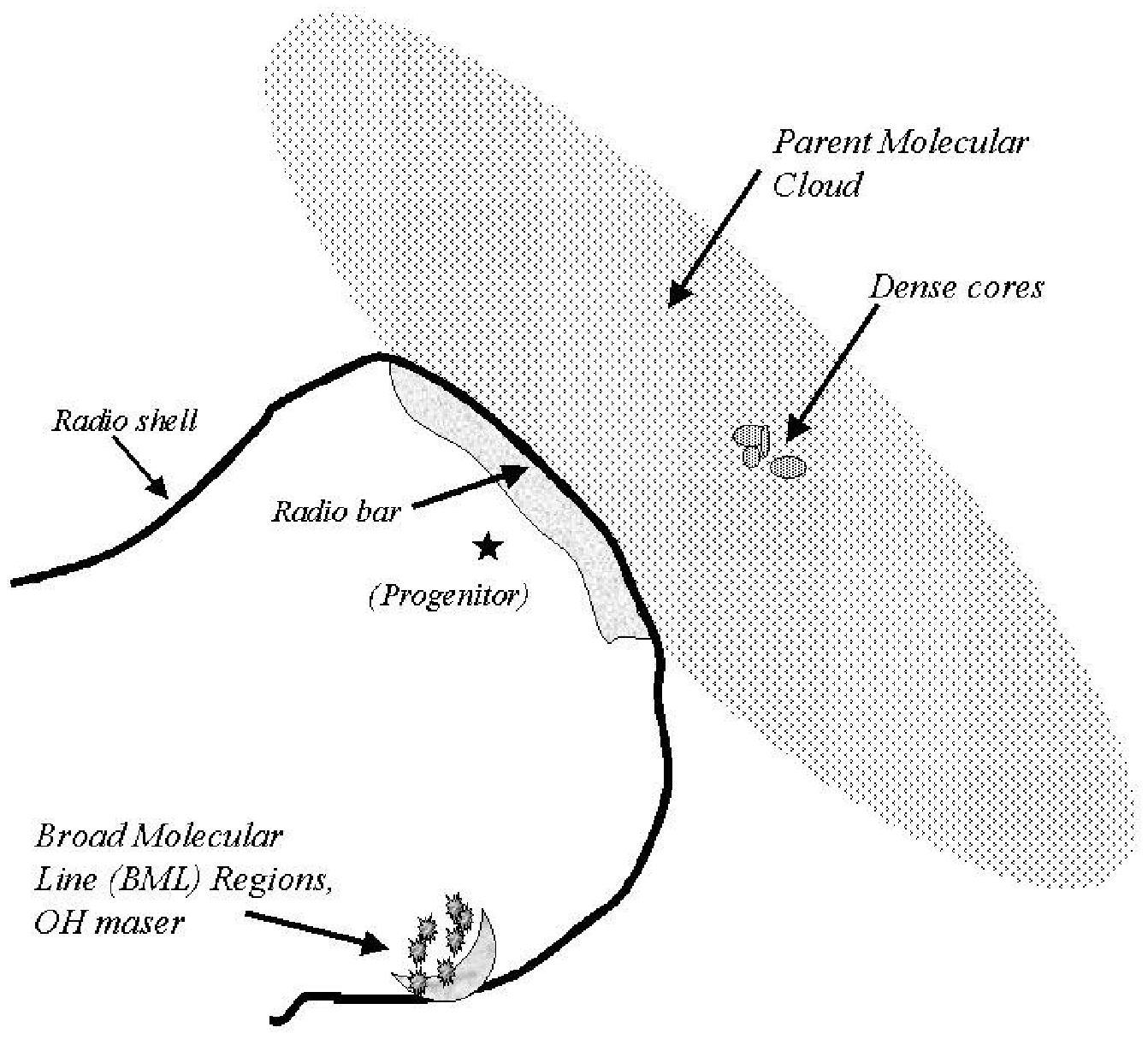}
\caption{Schematic illustration of 3C~391, showing the relative locations of
the radio shell, the radio bar, the parent molecular cloud, and the 
Broad Molecular Line (BML) regions. The possible location of the progenitor
(at the time of the supernova explosion) is indicated, as well as some
hypothetical dense cores in the parent cloud.
\label{cartoon}}
\end{figure}

\begin{figure}
\epsscale{.8}
\plotone{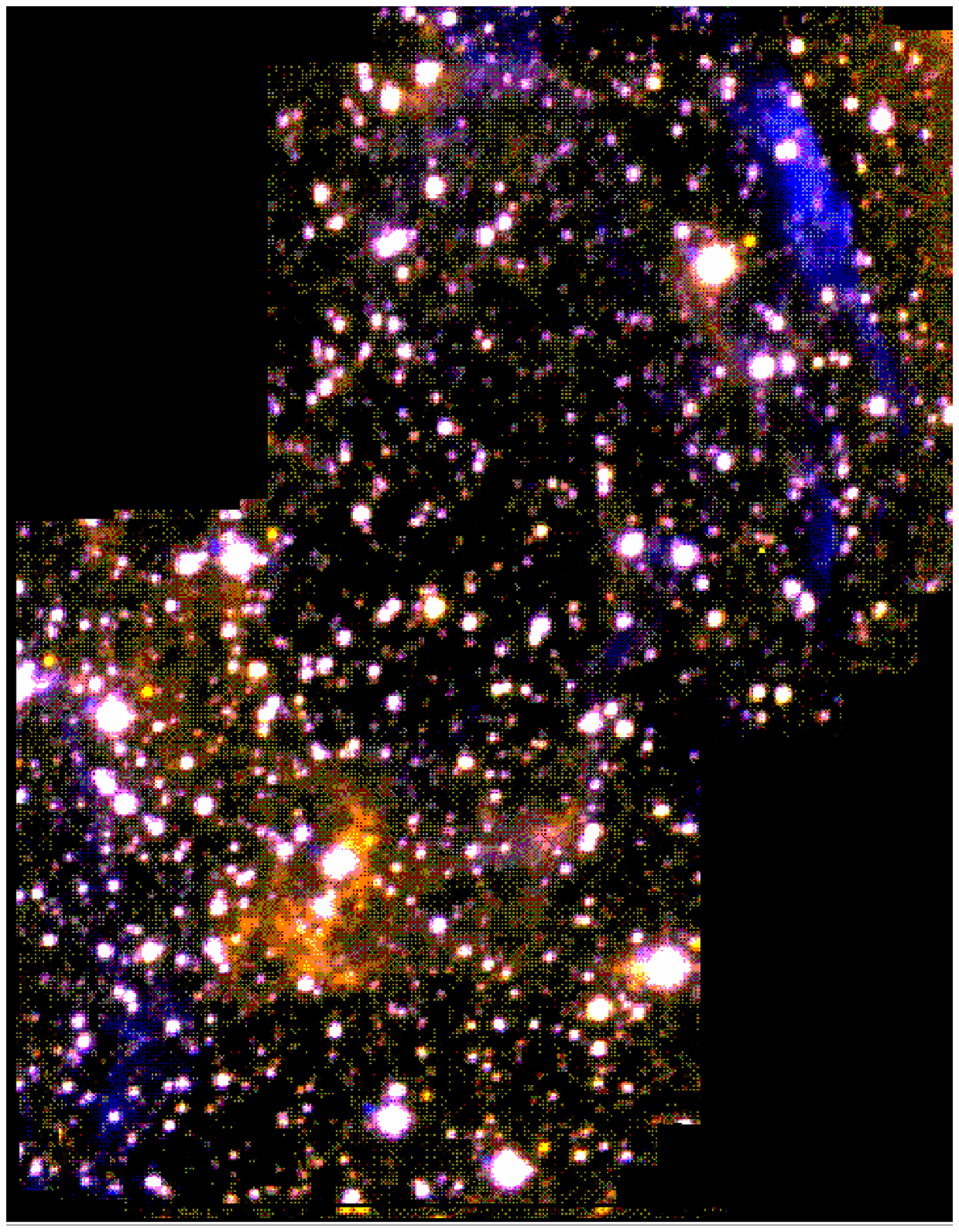}
\caption{Palomar 200$''$ color image of 3C~391. The H$_2$ 2.12 $\mu$m image is shown in red 
and green, and the [\protect\ion{Fe}{2}] 1.644 $\mu$m image is shown in blue. 
In the northwest, the [\protect\ion{Fe}{2}] emission
is bright at the location of brightest radio emission, which is right at the interface
between the remnant and the fact of the parent molecular cloud. The H$_2$ emission is only
bright in the southern part of the remnant, in a relatively coherent region that includes
the position of one of the OH masers. The distributions of H$_2$ and [\protect\ion{Fe}{2}] emission
are in stark contrast, demonstrating that the molecular and ionic shocks arise in 
completely different regions.
\label{paltrue}}
\end{figure}

The most straightforward result of our observations is that the molecular
and ionic emission comes from different regions. 
Figure~\ref{cartoon} is a schematic indicating the locations of the
different `geographic features' in 3C~391 to which we will refer in
this and the following sections. 
Figure~\ref{paltrue}
shows a false color image using the Palomar [\ion{Fe}{2}] and H$_2$ images.
The [\ion{Fe}{2}] emission is closely
associated with the bright radio bar in the northwestern part of the
remnant, with the [\ion{Fe}{2}] bar fitting precisely into an intensity
contour of the radio map. The other [\ion{Fe}{2}] filaments in the 
northwestern area are also associated with
radio ridges. The H$_2$ emission arises predominantly from a
clump in the southeastern portion of the image. The H$_2$ emission is
unfortunately centered near a bright star, and the seeing was not
optimal for our observations; however, the extended H$_2$ nebula can 
be clearly seen. On the angular scale of a few arcsec ($\sim 0.1$ pc), 
the molecular and ionic shocks are essentially exclusive of each other.

\begin{figure}
\epsscale{.5}
\plotone{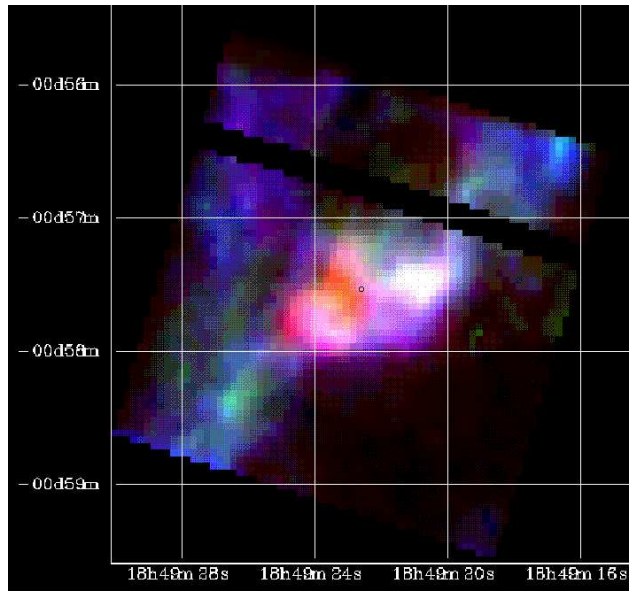}
\caption{ISOCAM color image of the southern portion of 3C~391. 
The location of this image relative to the rest of the remnant is shown in
Fig. 1. One of the ISOCAM columns was dead, leading to the
empty swath from upper left to middle right of this image.
The H$_2$ S(5) image is shown in red, the [\protect\ion{Fe}{2}] 5.5 $\mu$m
emission in green, and the [\protect\ion{Ne}{2}] image in blue. 
The H$_2$ emission arises predominantly from a small clump near the
center of the image. The ionic emission arises from filaments stretching 
from upper right to lower left and from the center to the upper left
of this image. These ionic filaments lie precisely along the radio shell
and an interior filament evident in the radio image.
\label{isotrue}}
\end{figure}

The {\it ISO} images further demonstrate differences between molecular
and ionic shocks. 
Figure~\ref{isotrue} shows a false color combination of three emission-line
images from the ISOCAM CVF data.
These mid-infrared images are significantly cleaner
than the near-infrared images because stars are so much fainter in the
mid-infrared, while the energy levels producing the emission lines are
generally closer to ground state and more easily excited in the dense
cooling layers behind the shock front. 
It is clear that the H$_2$ emission arises from a resolved region
near the center of the ISOCAM images. The [\ion{Fe}{2}], [\ion{Ne}{2}], and
[\ion{Ne}{3}] images are very similar to each other, but they are very
different from the H$_2$ images. The ionic lines follow the radio shell
and a radio filament that rises northward from the H$_2$ clump.
The ionic emission actually has a small `cavity' at the location of
the H$_2$ clump, as if part of the material that would have made the 
ionic shell is instead making the H$_2$ clump. This configuration is
not the most obvious one, for the ionic shocks are significantly
faster and should leave the dense, molecular clumps behind. 
The projected location of the H$_2$ clump
close to the edge of the remnant suggests that the H$_2$ clump
has only recently been shocked: given ionic shock speeds of
order 100 km~s$^{-1}$, and displacements between the molecular clump
and ionic shock less than 10$''$ (0.4 pc),
the shock front first encountered the molecular clump less than
4000 yr before the light we are observing was emitted. The age of
the remnant is thought to be only a few times greater than 4000 yr,
so when we say that the H$_2$ shock was recent, we only mean that it
was in the last $\sim 1/4$ of the lifetime of the remnant.

The new results presented in this paper complement previous results
from far- and mid-infrared spectroscopy \citep{rr00}. The spectroscopic
results showed that H$_2$ and ionic emission were both very bright and
both coming from the same general location in 3C~391. However, at
the angular resolution of the spectrometers that were used ({\it ISO}
LWS: 80$''$, SWS: $20''\times 33''$), regions with very different physical
properties were lumped together. The ISOCAM image clearly resolves
the ionic emission from the molecular emission toward 3C~391:BML.
The 3C~391:BML position was the only position (out of the sample
toward 3C~391, W~44, and W~28) for which a multiple far-infrared
molecular emission lines (CO, OH, and H$_2$O) were detected
\citep{rr98}, and it is the only position within the remnant 3C~391
toward which broad molecular lines were detected \citep{rr99}.
The Palomar image shows that the molecular emission is mostly
confined to the 3C~391:BML region, and that the ionic emission
is bright only in the northwest radio bar. The {\it ISO} SWS spectrum toward
the radio bar was the only one (out of the sample toward
3C~391, W~44, and W~28) that showed higher-excitation emission 
from [\ion{O}{4}]. Thus, while the spectroscopic results 
implied pre-shock regions of drastically different physical conditions,
the new imaging observations clearly demonstrate the separation
of these regions and prove that the blast wave is propagating
into a highly heterogeneous interstellar medium.

\section{Molecular shocks\label{molsec}}

The 3C~391 supernova remnant is near the surface of a molecular cloud,
with a bright radio bar that is very nearly tangent to that
surface \citep{wilner}.
One of the relatively surprising results of our infrared observations
of 3C~391 is that the brightest molecular emission does not arise from 
the interface with the parent molecular cloud. The millimeter-wave
observations led to a similar conclusion, because the CO and CS
emission lines near the radio bar have such narrow velocity widths
and low excitation that they look like quiescent, ambient gas. It would
be hard to hide shocked molecular gas from both the infrared and
millimeter-wave observations. Thus, it appears that the shocks at
the interface between the remnant and the molecular cloud are
nearly completely {\it dissociative}: the molecules are destroyed
and the dust grains partially vaporized (but not turned into
molecules). In the following subsections we will discuss the 
shock-excited molecular gas, and in the next section we will 
return to the ionic shocks and their relationship to the parent
molecular cloud.

\subsection{Comparison of near-infrared and millimeter-wave images}

\begin{figure}
\epsscale{.8}
\plotone{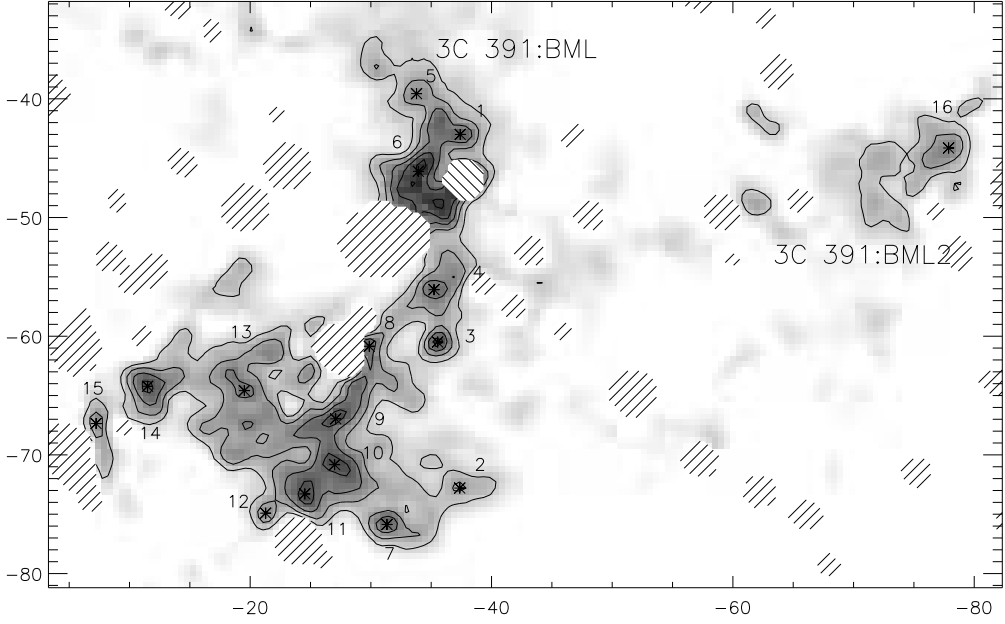}
\caption{Contour map of the continuum-subtracted H$_2$ image made
with PFIRCAM. 
Contour levels begin at 0.25 and have an interval of 0.5,
in units of $10^{-4}$ erg~s$^{-1}$~cm$^{-2}$~sr$^{-1}$.
The coordinates are offsets, in arcsec, from 
$18^h49^m24.8^s  -00^\circ 56^\prime 31.1^{\prime\prime}$ (J2000), 
which is the same nominal position used in the millimeter-wave
observations, allowing direct comparison to Figs. 5 and 6 of
Reach \& Rho (1999).
The H$_2$ clumps are labeled with their entry number in Tab.~\ref{h2peaks}. 
The OH maser is not
located at the center of the H$_2$ nebula, but rather it is the minor but 
very well-defined H$_2$ peak number 3.
Diagonal hatching indicates positions containing stars so bright that the 
continuum subtraction was inaccurate. The small patch with diagonal hatching
in the `upper-left to lower-right' direction, near ($-3''$,$23''$) 
was masked because it is contaminated by a ghost image artifact of the
bright star near ($8''$,$17''$).
\label{h2clump}}
\end{figure}

The location, size, and shape of the region
that produces the mid-infrared H$_2$ lines are
the same as the region that produces 
the broad millimeter-wave lines in 3C~391:BML. 
Figure~\ref{h2clump} shows a
contour map of the near-infrared H$_2$ emission with the continuum (stars)
subtracted. 
For ease of quantitative comparison, the coordinates
are expressed as offsets from the same position used to offset the
IRAM 30-m telescope. Comparing to Fig. 5 of Reach \& Rho (1999), 
all of the pointings of
the IRAM 30-m telescope that showed wide CO(2-1) emission
also show bright H$_2$ emission. 
The millimeter-wave
observations were sensitive both to shock-excited gas and cold gas,
and they showed a peak in both the broad-line and narrow-line gas
emission. The broad-line emission follows the H$_2$ emission,
at the $10''$ angular resolution of the CO($2\rightarrow 1$) data.
The narrow-line gas is located immediately to the southeast of the
broad-line region, centered at relative offsets $-10'',-85''$, just
off the lower left corner of Fig.~\ref{h2clump} in a region with no
H$_2$ emission. This is consistent with the broad-line gas being
shock-excited, while the narrow-line gas is unshocked, ambient gas.
Further, since it appears that the hot H$_2$ emission (seen in
the ISOCAM and PFIRCAM images) arises from the same 
location as the broad-line millimeter emission, the shock must
have heated the molecules without destroying them.

\subsection{Morphology of the shocked H$_2$}

Looking at the near-infrared image in more detail, it is evident that
the shocked H$_2$ emission breaks up into a relatively
structured, clumpy distribution. 
In Figure~\ref{h2clump}, the emitting region is very structured on the angular
scale of the previous spectroscopy, and appears to show clumping 
down to the limit of the atmospheric seeing during our observation.
We will refer to the entire H$_2$-emitting nebula in Fig.~\ref{h2clump} 
as 3C~391:BML (which is coincident with the shocked
broad-molecular-line region in Figs. 5 and 6 of \citet{rr99}).
The main body of this nebula is irregularly shaped, 
$35^{\prime\prime}\times 45^{\prime\prime}$ in size,
with a geometric center 
around $18^h49^m23.1^s -00^\circ 57^\prime 33^{\prime\prime}$ (J2000).
The individual clumps in the nebula are labeled in Figure~\ref{h2clump} from
1 through 16, and we will refer, e.g. to `clump 7' as `3C~391:BMLH2 7.'

\begin{figure}
\epsscale{1}
\plotone{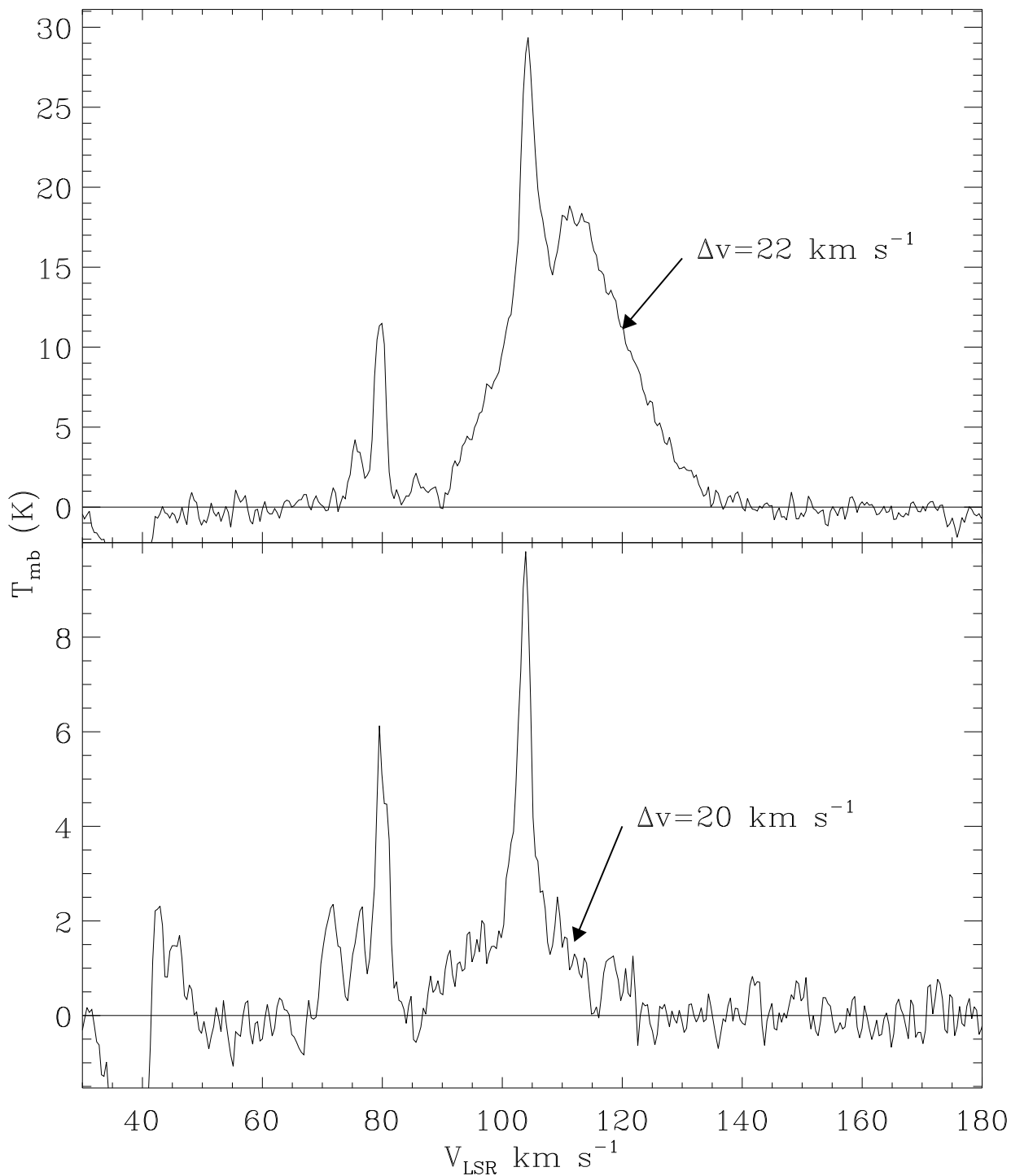}
\caption{Spectra of the CO($2\rightarrow 1$) line toward the main body of
3C~391:BML (top) and the small clump
3C~391:BMLH2 16 (bottom). The broad-line emission toward 3C~391:BMLH2 16 is 10 times weaker
than toward 3C~391:BML, at least for the brightnesses averaged over the 10$''$ beam 
of the IRAM 30-m telescope. The full widths at half maximum intensity of the
wide components of the spectral lines are labeled with arrows.
\label{h2newco}}
\end{figure}

To verify that the faint emission is real and correlated with molecular
emission, we examined a small region of H$_2$ emission, 3C~391:BMLH2 16.
The ISOCAM CVF spectrum of this clump reveals bright mid-infrared H$_2$ lines
together with ionic lines. In the color ISOCAM image,
3C~391:BMLH2 16 is along an [\ion{Fe}{2}] filament but also
a peak in the H$_2$ image; therefore, it appears white in 
Figure~\ref{isotrue}. In Fig.~\ref{paltrue}, 3C~391:BMLH2 16 appears more blue
than the main body of 3C~391:BML, because of the 
brighter [\ion{Fe}{2}] from the former.
3C~391:BMLH2 16 is outside the region where broad molecular line emission
was previously noticed, so we derived the CO($2\rightarrow 1$) spectrum 
from the previously published observations \citep{rr99}. 
Figure~\ref{h2newco} shows that there is indeed broad-line emission
from 3C~391:BMLH2 16. The broad-line emission is 10 times fainter
than brighter parts of 3C~391:BML, but its line width is comparable. 
Considering that the H$_2$ emission from 3C~391:BMLH2 16 is fainter and 
the clump is diluted in the $10''$ beam of the IRAM 30-m telescope (used 
to make the CO spectra), 
it is not surprising that the broad-line emission is fainter than for
the main body of 3C~391:BML. The comparable CO line widths suggest that the
shocks being driven into the two molecular clouds are similar.
The centroids of the broad molecular lines are somewhat different,
with the main body of 3C~391:BML redshifted by approximately 6 km~s$^{-1}$ with respect
to 3C~391:BMLH2 16; the redshift is most likely due to a somewhat different
angle of the shock front with respect to the line of sight. 
The ability of the H$_2$ image to predict the presence of broad molecular
lines, as confirmed by the detection of 3C~391:BMLH2 16, shows that 
shocked H$_2$ emission is a practical tracer of dense molecular shocks.

Table~\ref{h2peaks} lists the 16 main peaks of H$_2$ emission, together
with their central intensities and major and minor axes. The
clumps are resolved in at least one dimension, with a typical
size of order 0.1 pc. The clumps in our list account for most but not all
of the H$_2$ emission. At the angular resolution of our observations,
the morphology of the H$_2$ emission is different from that 
of shocks into lower-density gas: the H$_2$ is clumpy, while the
[\ion{Fe}{2}] emission is filamentary. For most supernova remnants,
the optical emission is filamentary. The special morphology of the
H$_2$ emission reflects the distinct physical properties of the 
pre-shock gas. While most supernova shocks seen in optical and radio images
propagate into gas with
density $n_0<10$ cm$^{-3}$, the shocks in 3C~391:BML are propagating
into regions of much higher pre-shock density, with $n_0>10^4$ cm$^{-3}$
\citep{rr00}. 

\begin{table}[tbh]
\caption[]{H$_2$ Clumps in the 3C~391:BML nebula}\label{h2peaks} 
\begin{flushleft} 
\begin{tabular}{lcccccc} 
\hline
clump & RA,Dec & Major & Minor & $10^4 I_{peak}$ & $L(2.12\,\mu{\rm m})$ & $M({\rm H}_2)$ \\
      & (J2000) & (pc) & (pc)  & (erg s$^{-1}$ cm$^{-2}$ sr$^{-1}$) &
      	 ($L_\odot$) & ($M_\odot$) \\
\hline
 1 &  18 49 22.4  -00 57 14.7 & 0.086 & 0.065 & 0.82 & 0.06 & 0.79 \\
 2 &  18 49 22.5  -00 57 44.3 & 0.108 & 0.065 & 0.86 & 0.08 & 1.03 \\
 3 &  18 49 22.6  -00 57 32.0 & 0.065 & 0.043 & 1.54 & 0.06 & 0.74 \\
 4 &  18 49 22.6  -00 57 27.6 & 0.151 & 0.065 & 0.66 & 0.08 & 1.11 \\
 5 &  18 49 22.6  -00 57 11.2 & 0.086 & 0.086 & 0.55 & 0.05 & 0.70 \\
 6 &  18 49 22.6  -00 57 17.7 & 0.108 & 0.086 & 0.93 & 0.11 & 1.48 \\
 7 &  18 49 22.9  -00 57 47.1 & 0.129 & 0.108 & 0.95 & 0.17 & 2.27 \\
 8 &  18 49 22.9  -00 57 32.2 & 0.129 & 0.108 & 1.51 & 0.27 & 3.64 \\
 9 &  18 49 23.1  -00 57 38.2 & 0.129 & 0.065 & 0.83 & 0.09 & 1.20 \\
10 &  18 49 23.1  -00 57 42.0 & 0.129 & 0.086 & 1.82 & 0.26 & 3.49 \\
11 &  18 49 23.3  -00 57 44.4 & 0.108 & 0.108 & 1.87 & 0.28 & 3.74 \\
12 &  18 49 23.5  -00 57 45.9 & 0.108 & 0.086 & 0.72 & 0.09 & 1.15 \\
13 &  18 49 23.6  -00 57 35.7 & 0.151 & 0.086 & 0.57 & 0.10 & 1.28 \\
14 &  18 49 24.1  -00 57 35.1 & 0.172 & 0.172 & 0.70 & 0.27 & 3.59 \\
15 &  18 49 24.4  -00 57 38.1 & 0.129 & 0.065 & 1.01 & 0.11 & 1.45 \\
16 &  18 49 19.7  -00 57 16.9 & 0.127 & 0.086 & 1.62 & 0.23 & 3.05 \\
\hline
\end{tabular} 
\end{flushleft} 
\end{table}

The clumps in our H$_2$ image are probably are dense cores that have survived
the passage of the blast wave. 
The clump masses were estimated from the observed brightness of the 2.12 $\mu$m
line using a 10-level model for the H$_2$ molecule over a range of
densities and with kinetic temperature 1300 K (derived in the next
section); the results can be approximated by
\begin{equation}
M \simeq 0.8 \left(\frac{10^5 {\rm~cm}^{-3}}{n({\rm H}_2)}\right) 
\left(\frac{I(2.12 \mu{\rm m})}{10^{-3} {\rm erg~s}^{-1}{\rm ~cm}^{-2}{\rm ~sr}^{-1}}\right) 
\left(\frac{\theta}{1''}\right)^2 M_\odot
\end{equation}
where $\theta$ is the clump angular diameter at the 9 kpc distance assumed
for 3C~391.
The total clump mass can be higher than our
calculated mass because the near-infrared line only detects the excited 
H$_2$ and not the dense, cold H$_2$ in the centers of the clumps.
Analysis of the excitation of broad-line CS emission implied that the
density of the mm-wave emitting region is $n\sim 2\times 10^5$ cm$^{-3}$
\citep{rr99}. A clump with $2''$ diameter would therefore have a
mass of 0.6 $M_\odot$, comparable to the properties inferred from the
H$_2$ observations, suggesting that we have resolved the mm-wave 
emitting regions.
The clumps in Table~\ref{h2clump} would be gravitationally bound
if their internal turbulent motions were less than 
$\sim 1$ km~s$^{-1}$, which is typical for quiescent gas.
Thus the pre-shock clumps were likely self-gravitating.

\subsection{Implications for star formation}

It appears that the shock front passing through 3C~391:BML has
uncovered a set of pre-stellar cores. 
We say that the cores are `pre-stellar' because, even though they appear
centrally condensed in the H$_2$ image, there are no stars 
at the core locations: in the 2.19 $\mu$m continuum image, there are
no stars down to K magnitude 15 (1 mJy). And in the mid-infrared 
(12--18 $\mu$m) continuum image, there are no stars down to 10 mJy. 
Embedded stars would need very high K-band extinctions but low 
12--18 $\mu$m flux.
Thus, the cores probably do not harbor young stars.

The fate of these prestellar cores, after having been perturbed by the
supernova shock, is worth considering in detail, because many
star-forming molecular clouds will suffer internal or
surface supernova explosions. The outer layers of the dense cores will be
stripped by the ram pressure of the shock. Balancing the ram-pressure
force, $F_{ram} \sim p_{ram} \pi R^2$,
versus the gravitational force binding a thick layer with radius $R$ from
the center,$F_{grav} \sim G M^2 / R^2$,
we find a stripping radius of
\begin{equation}
R \sim 10^{17} \left(\frac{r_{10}^3}{E_{51}}\right)^{1/4} 
\left(\frac{M}{M_\odot}\right)^{1/2}\,{\rm cm},
\end{equation}
where $E_{51}$ is the supernova energy in units of $10^{51}$ erg, 
$r_{10}$ is radius of the (assumed adiabatic) remnant in units of 10 pc, 
and $M$ is the mass of the clump.
The stripped radius is comparable to the observed size of the clumps,
which is consistent with the hypothesis that the clumps are gravitationally
bound cores stripped of their outer layers. 
The stripped radii are smaller than the apparent outer edges of 
unshocked pre-stellar cores inferred from mid-infrared extinction 
measurements \citep{bacmann}.
Clumps encountered earlier in
the remnant evolution would be even more stripped or completely distorted,
to become a shocked filament rather than retaining integrity as a core.
For the surviving clumps, with the outer layers gone, the accretion onto the 
central core will slow, perhaps stunting the development of the star. 

If a core has a radial density profile proportional to $r^{-2}$,
then the observed masses ($\sim M_\odot$) and radii ($\sim 10^{17}$ cm)
suggest densities of order $5\times 10^{4} r_{17}^{-2}$ cm$^{-3}$.
The lifetime estimates for pre-stellar cores of this density are 
of order $10^{5.6}$ yr \citep{andre}, longer than the remnant age
and consistent with the idea that the several observed cores were in 
an intermediate, pre-stellar phase at the time of the supernova.
If the pre-shock cores were already gravitationally bound, 
they may have been already collapsing in the center, 
with the outer parts supported by turbulence and magnetic field. The magnetic field 
and angular momentum of the outer layers would be swept away,
so the pre-stellar core would no longer be as coupled to its surroundings 
as it would be if it had not been shocked. The net effect on the
star formation potential of the shocked cores could be rather 
subtle, as a very dense central core may be relatively immune to
the effects of a supernova shock. This topic deserves theoretical treatment,
as it concerns the connection between environmental effects and
star formation in the not-uncommon situation of molecular cores within
10 pc of a supernova explosion.

One of the H$_2$ peaks, 3C~391:BMLH2 3, is precisely at the position
of one of the two OH 1720 MHz masers in 3C~391 \citep{Frail96}. 
The astrometry was determined using a set of 12
stars from the 2MASS catalog, and we checked for local distortion by
precisely measuring the offset from the bright star near the center of
3C~391:BML. The astrometric accuracy and the radio-infrared offset
are both smaller than $0.3''$. The H$_2$ peak associated with the OH
maser appears to be resolved into a round clump with a FWHM of $2.5''$.
(In the 1\% H$_2$ filter, the clump brightness is equivalent to a magnitude 15.6 star,
while in a broad K$_s$ filter, the clump would be equivalent to magnitude 18.5.)
This `maser clump' accounts for only $\sim 3$\% of the entire H$_2$ luminosity of
3C~391:BML. Thus, while the OH 1720 MHz
masers are `signposts' of molecular shock fronts \citep{mitchell}, 
the OH masers are not one-to-one tracers of molecular shocks.
Not all molecular shocks give rise to OH masers---other 
clumps that appear similar to 3C~391:BMLH2 3 (the `maser clump')
do not have bright, mased OH emission. Our observations may offer some
clues to the nature of the OH 1720 MHz masers. If our interpretation of
the H$_2$ clumps is correct, then the 1720 MHz masers may be features of 
protostars or pre-stellar cores that were partially uncovered by the shocks. 
The precise combination of
physical conditions required to make a 1720 MHz OH maser 
may occur in specific layer (or region) in the atmospheres
of pre-stellar, shocked cores; indeed,
OH masers are known to be signposts of star formation \citep{gaume87}.

\subsection{Excitation of H$_2$}

\begin{figure}
\epsscale{1}
\plotone{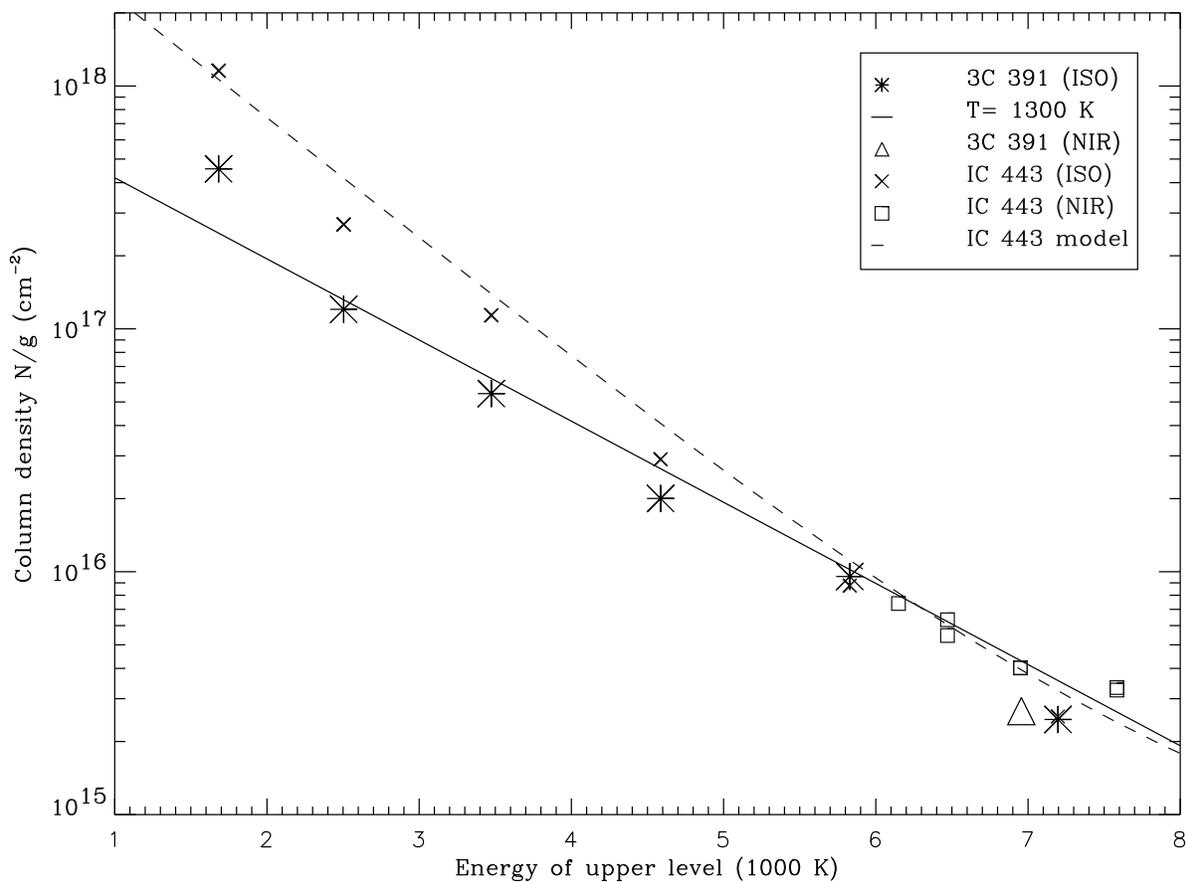}
\caption{H$_2$ excitation diagram for 3C~391:BML. The asterisks 
show the column densities per unit degeneracy ($g$) for the upper 
level of each of the mid-infrared lines observed with {\it ISO}.
The triangle is based on the 
brightness of the near-infrared 2.12 $\mu$m line for a typical clump in
Fig.~\ref{h2clump} or Tab.~\ref{h2peaks}. The straight, solid line
shows the populations for H$_2$ in LTE at 1300~K. 
The dashed line shows the multi-temperature fit 
derived from near- to mid-infrared observations of
IC~443 (Reach \& Rho 2000).
\label{h2exc}}
\end{figure}

The 6 lines of H$_2$ that were detected in the ISOCAM data span a range of upper 
level energies lying 1682 to 7197 K above ground, so they sample warm,
collisionally-excited gas recently heated by shocks. Figure~\ref{h2exc}
shows the upper-level column densities for each of the observed lines,
after correcting for extinction.
The populations of levels 2000 K and higher above the ground state 
are well approximated by local thermodynamic equilibrium (LTE) at 1300~K
with a warm H$_2$ column density of $1.6\times 10^{20}$ cm$^{-2}$.
Some support for our
extinction estimate is that the S(3) line fits nicely in Figure~\ref{h2exc};
this line is at 9.66 $\mu$m and is extinguished by the 
silicate feature of the interstellar extinction curve.
Regarding the ortho-to-para ratio, 
the ISOCAM spectrum includes 3 transitions of para-H$_2$ and 3 lines of ortho-H$_2$
(asterisks in Fig.~\ref{h2exc}). Both ortho and para transitions fall along the same trend
in Fig.~\ref{h2exc}, so the ortho-to-para ratio is close to 3.

The excitation of the near-infrared line is somewhat less than predicted
by a single-temperature LTE model. In LTE, a model that matches the 
brightness of the 0-0 S(7) line would accurately
predict the brightness of the 1-0 S(1) line, because their
upper energy levels are very close. However, the critical density for the 
upper level of the 1-0 S(1) line is $\sim 6\times 10^6$ cm$^{-3}$, an
order of magnitude larger than that of the 0-0 S(7) line, so the 1-0 S(1)
line may be subthermally excited. We calculated the excitation
of the 10 lowest levels of H$_2$ using the excitation rate coefficients for
H$_2$-H$_2$ collisions as approximated by \citet{DRD}; the results
show that gas with $T\sim 1300$ K and $n({\rm H}_2)\sim 10^5$ cm$^{-3}$
could explain the observed excitation of the levels between 2000 and 8000 K
above ground, including the levels that produce the near-infrared line.

A single-temperature model cannot explain all
the observed line brightnesses: the models miss the lowest-energy
line by a factor of 2. Therefore, the 
single-temperature models under-predict the H$_2$ column density.
For the
bright, well-observed remnant IC~443, a two-temperature model could adequately
explain the mid-infrared lines \citep{cesarsky443}. 
Figure~\ref{h2exc} compares the excitation model derived from near- to 
mid-infrared observations of IC~443 \citep{rho443} and our 3C~391 data.
The IC~443 model did not have to be rescaled for this comparison.
For both shocks, the column density inferred
from observations of the near-infrared lines alone is smaller than
that inferred from the lower-energy mid-infrared lines. The actual excitation
is non-LTE, and it depends both on the volume density of the emitting
region and possibly on the time that the gas has cooled \citep{cesarsky443}.
We used the two-temperature excitation model to calculate the clump column
densities in Table~\ref{h2peaks}.

\section{Ionic shocks}

\begin{figure}
\epsscale{1}
\plotone{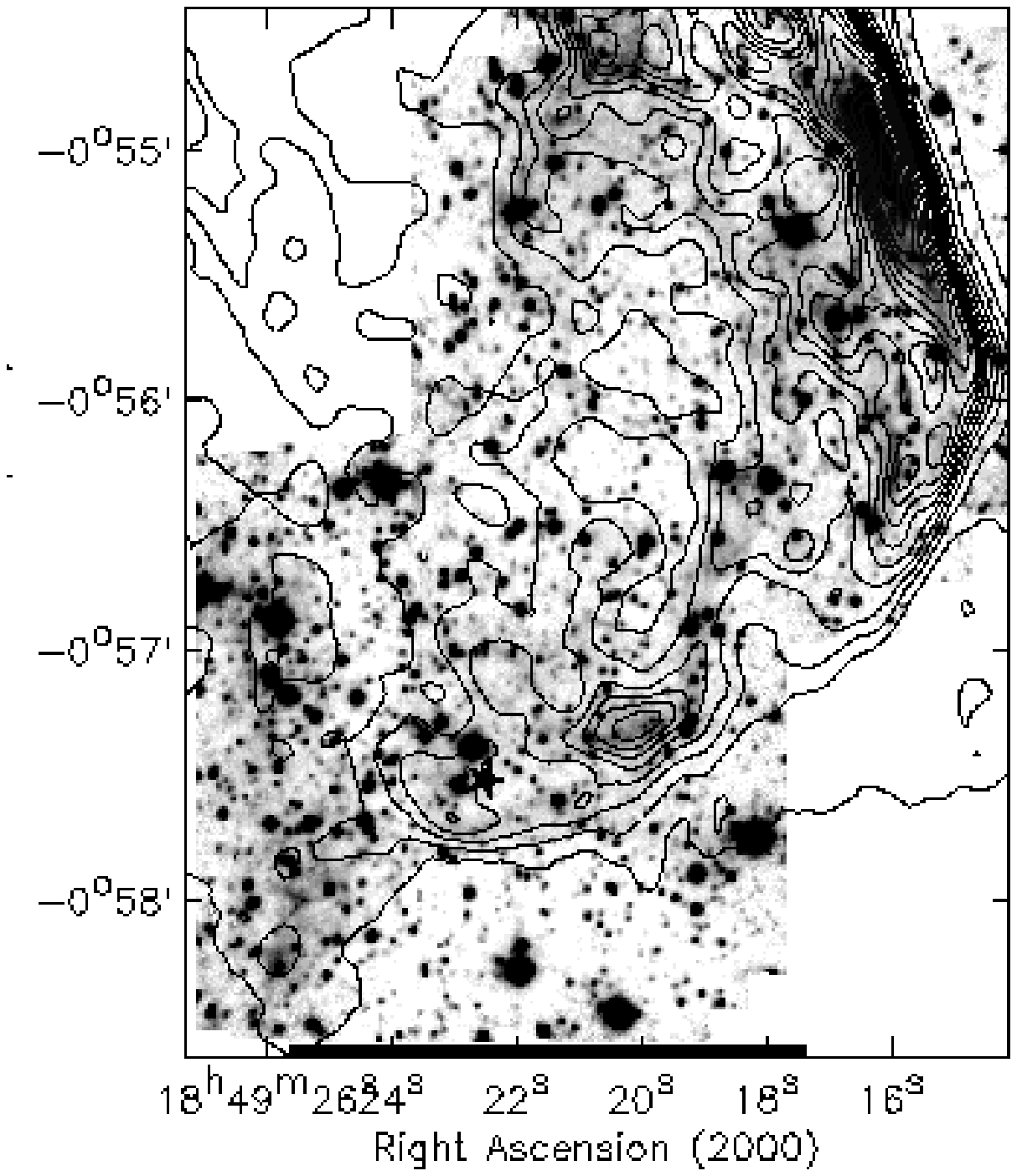}
\caption{[\protect\ion{Fe}{2}] image (greyscale) with radio continuum contours overlaid.
The first 10 radio contours (black) are linearly spaced with an interval of 0.025 Jy/beam, 
then the upper 7 radio contours (white) are linearly spaced with an interval of 0.05 Jy/beam.
The radio image has a $6''$ cleaned beam; it was projected onto the [\protect\ion{Fe}{2}] image,
which has $0.5''$ pixels, with $1.6''$ seeing.
The radio bar in the northwest is much brighter than the rest of the remnant,
and the shape and size of the bar is the same in [\protect\ion{Fe}{2}] and radio continuum.
The faint [\protect\ion{Fe}{2}] over the rest of the field is correlated with the radio
continuum in detail. This includes arcs parallel to, but just inward from, the radio 
bar, and the western border of the remnant, which is delineated by a narrow
filament that runs roughly
north-south at R.A. $18^h49^m19^s$. 
This filament ends in a knot of enhanced [\protect\ion{Fe}{2}] and radio emission
at $18^h49^m20.5^s-0^\circ 57' 20''$. Among other detailed correspondences
are the [\protect\ion{Fe}{2}] and radio enhancements just east of the radio
bar, around $18^h49^m20.8^s-0^\circ 54' 43''$.
\label{fe_radio}}
\end{figure}

The ionic shocks are traced, in our data, by emission from 
[\ion{Fe}{2}], [\ion{Ne}{2}], and [\ion{Ne}{3}], which follows
the shell-like radio emission rather closely. Figure~\ref{fe_radio} 
the Palomar [\ion{Fe}{2}] image with the resampled radio image
overlaid as contours. 
The bright [\ion{Fe}{2}] in the northwest agrees in detail with the size and
location of the bright radio bar, and the other features, though fainter,
also seem to agree in detail.

The morphology of the [Ne] emission also seems to follow the radio
emission in some detail. From the CVF spectra, it appears that the 12--18 $\mu$m filter
image, after subtracting the reference spectrum, is dominated by the 
[\ion{Ne}{2}] 12.8 $\mu$m and  [\ion{Ne}{3}] 15.5 $\mu$m
lines. Thus the diffuse emission in Figure~\ref{figlw3} is effectively
a map of [Ne], at least for the part associated with the remnant.
(Fig.~\ref{figlw3} also
contains aromatic hydrocarbon emission from unrelated gas, but for the structure 
related to 3C~391:BML, the spectra in Figs.~\ref{clumpspec} and~\ref{ionspec} show that
there is no aromatic hydrocarbon emission or continuum from the remnant.)
From this figure we see that the radio bar is replicated in [Ne].
The radio ridge at the easternmost
part of the remnant ($18^h49^m36.6^s-0^\circ 55' 23''$) is also present in
the [Ne] image, but with an apparent shift such that the [Ne] is partly
`outside' of the radio shell. Care is needed in interpreting wide-band
images such as Figure~\ref{figlw3}, because the spectrum can vary significantly
from place to place. The disagreement between the radio and 12--18 $\mu$m image may
in this case be due to a contribution from dust grains, as is suggested
by comparing to the 10.8--11.9 $\mu$m image (see \S 6).

The agreements and differences between the [Ne], [Fe], and radio images
are the result of combined effects of excitation, grain destruction, and
magnetic field. 
In general, the [\ion{Fe}{2}] image
appears better-correlated with the radio emission than the [Ne] image.
The close agreement of [Fe] and radio emission suggests that
the shocks that are most able to accelerate cosmic rays are also the ones
capable of destroying dust grains. The [Ne] emitting regions may be shocks
that are less efficient at grain destruction because of higher pre-shock density,
leading to lower shock velocity; grain destruction is very sensitive to the
shock velocity \citep{jones96}. The differences between [Ne] and [Fe] could
also be due to excitation differences, because the [Ne] lines we observed are
among levels that are closer to the ground state than the levels that
produce the near-infrared [\ion{Fe}{2}] line.

The [\ion{Fe}{2}] 1.64 $\mu$m emission from the radio bar is bright,
suggesting a hot postshock region with a significant fraction of grains
destroyed. The [\ion{Fe}{2}] 
26 $\mu$m line was observed toward the northern part of 3C~391
using {\it ISO} SWS, in a $14''\times 27''$ aperture \citep{rr00}. The SWS pointing was
centered on an earlier LWS pointing that was part of a strip map crossing the northern
edge of the remnant. It appears that the SWS pointing called 3C~391:radio does 
not sample the brightest part of the radio bar, where our new Palomar image
reveals a highly concentrated bar of emission. Instead, the SWS observation
sampled a region of much fainter [\ion{Fe}{2}] emission, $<0.2\times 10^{-4}$.
Comparing the near-infrared limit to the observed brightness of the 
26 $\mu$m line, using a 13-level excitation model for the Fe$^+$ ion,
we can conclude that the emitting region has a temperature less 
than 3000 K, which 
is consistent with the brightness of other lines observed behind the 
J-shocks into moderately dense gas.

The pre-shock density for the radio bar was likely comparable to that 
of the molecular cloud. 
Comparing the [\ion{O}{1}] brightness to J-shock models, a moderate
density $\sim 10^3$ was inferred \citep{rr96}.
The brightness and morphology to the radio bar of 3C~391 is very
similar to the northeast rim of IC~443, which has 
more complete spectroscopic observations.
The surface brightness and line ratios imply a pre-shock density or 
order $10 < n_0 < 10^3$ cm$^{-3}$ \citep{rho443}. Such densities are
actually comparable to the average densities of giant molecular clouds such as
are traced by rotational CO lines at mm-wavelengths. 
For example, the Rosette
molecular cloud \citep{williams95} and molecular clouds in the inner galaxy
\citep{simon01} were found to be composed of CO-emitting regions with densities
in the range $30 < n_0 < 10^3$ cm$^{-3}$ embedded in even lower density gas.
Thus the `ionic' shocks are plausibly shocks into the
CO-emitting pre-shock regions that traditionally define molecular clouds.

So far, we have discussed only ionic lines from the shocks in the
northwestern part of the remnant, arguing that these shocks are completely
dissociative. The excitation of the ionic lines, the shock
velocity inferred from the width of the [\ion{O}{1}] 63 $\mu$m line
\citep{rr00},
and the presence of the molecular cloud surface so nearly
tangent to the northwestern radio bar \citep{wilner} all
show that the pre-shock gas was mostly molecular. 
Furthermore, there is some weak emission from molecular
gas in the northwestern region: the H$_2$ S(3) line was detected
using the {\it ISO} SWS, at a brightness of 
$3\times 10^{-5}$ erg cm$^{-2}$ s$^{-1}$ sr$^{-1}$ \citep{rr00}. 
This emission is much weaker than that detected toward 3C~391:BML.
The H$_2$S(9), which arises from a much higher energy level than
the S(3) line, was not detected from the northwest; nor is there
1--0 S(1) line emission in the Palomar image. 
Thus, we interpret the weak, low-excitation H$_2$ emission 
from the northwestern part of 3C~391 as molecules that have {\it reformed}
behind a dissociative shock. For such a shock, the molecules reform after 
the gas has recombined and cooled, and the bulk of the H$_2$ 
emission arises in a layer with relatively low temperature
\citep{HM89}. In contrast, for a non-dissociative shock, the 
emission can arise from regions where the H$_2$ molecules reach 
significantly higher excitation, with energy levels potentially 
approaching the dissociation energy \citep{DRD}. 
The H$_2$ emission arises from molecules that were dissociated
in the shock and subsequently reformed on the surfaces of surviving
grains. As a corollary, some of the grains must survive the shock; this
is not inconsistent with the observed lack of mid-infrared aromatic features 
or continuum from the shock fronts, because larger grains (which do not produce 
mid-infrared features or continuum, but do dominate the mass and far-infrared 
emission of interstellar dust) may survive the shocks.
The timescale for reformation of H$_2$ molecules in the physical conditions of 
the 3C~391 ionic shocks is $\sim ~10^5$ yr, somewhat longer than the remnant
age; thus, H$_2$ reformation is probably only partial. 
One prediction
of this interpretation is that the velocity widths of the 
H$_2$ S(3) lines will
be $\sim 100$ km~s$^{-1}$ in the northwest and $\sim 25$ km~s$^{-1}$
in the 3C~391:BML region.

\section{Dust Destruction}

The lack of any significant continuum emission or solid-state or macromolecular
spectral features in the mid-infrared spectrum suggests that the shocks have
destroyed some of the grains.
The mid-infrared emission of unshocked interstellar
gas is dominated by solid-state features at 6.2, 7.7, 8.6, 11.3, and 12.6 $\mu$m
due to aromatic hydrocarbons; these features, combined, produce
the emission that was detected almost everywhere by {\it IRAS} \citep{pahref}.
These features are all clearly evident
in the ISOCAM CVF spectrum of unshocked gas in Figure~\ref{cvfspec}. 
However, the {\it shocked} gas shows no solid-state features:
both the molecular shock (Fig.~\ref{clumpspec}) and the ionic shock
(Fig.~\ref{ionspec}) spectra are devoid of aromatic features,
despite bright features from the pre-shock and unrelated gas
(Fig.~\ref{cvfspec}).
Further, the shocked emission shows no significant continuum rise at long wavelengths. 
This rise is normally seen in interstellar dust spectra, including those
of quiescent regions like reflection nebulae \citep{n7023}.
This mid-infrared rise is required to explain 
the bright 25 $\mu$m emission from the interstellar medium, as also detected
by {\it IRAS}, because there are no bright spectral solid-state
features at longer wavelengths. 
In H~II regions, the aromatic features are weak or absent, while the mid-infrared
continuum is relatively bright \citep{m17}. Thus, the mid-infrared emission from
the interstellar medium can be considered as due to two types of particles:
very small particles or macromolecules of aromatic hydrocarbons that
are the carriers of the 6.2--12.6 $\mu$m features, and very small solid dust
particles that are carriers of the mid-infrared continuum \citep{desboul}.
The lack of aromatic features or continuum in the mid-infrared spectrum of 3C~391
suggests that {\it both} carriers are destroyed.

\begin{figure}
\plotone{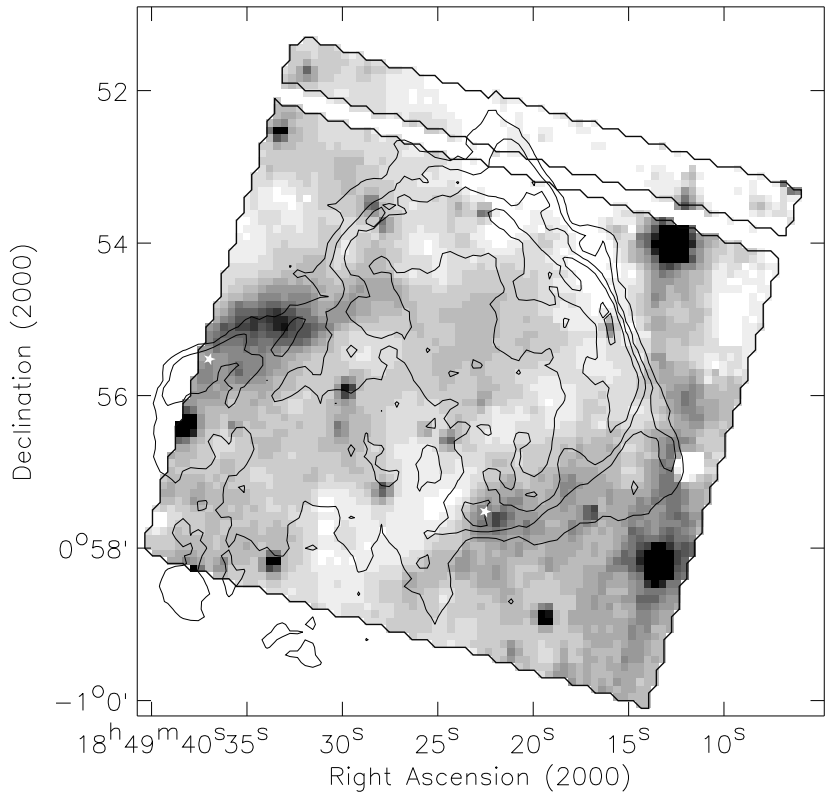}
\caption{ISOCAM LW8 (10.8--11.9 $\mu$m) filter image of 3C~391. The radio
contours (same as in Fig. 1) are overlaid for comparison. 
There is no LW8 emission following the radio shell, but there is a thick,
`shadow' shell that runs along the interior of the shell.
\label{figlw8}}
\end{figure}

To look more carefully for aromatic hydrocarbons, we made a small image of 3C~391
with ISOCAM in the LW8 filter (10.8--11.9 $\mu$m). This filter contains no 
significant spectral lines of shocked gas (see Fig.~\ref{clumpspec}), but it
is centered on the bright aromatic feature at 11.3 $\mu$m. Thus emission in this
filter traces the distribution of aromatic hydrocarbon particles. Figure~\ref{figlw8}
shows the 10.8--11.9 $\mu$m image. The supernova remnant is not evident in this image.
The total contrast in the 10.8--11.9 $\mu$m image is small; after subtracting 
105.8 MJy~sr$^{-1}$
of zodiacal emission appropriate for this filter and observing location
and date, the range is 29--40 MJy~sr$^{-1}$ for the diffuse emission in
the region observed by the ISOCAM CVF. The choice of CVF reference spectrum
(\S 2.1) location is relatively clean of structure and typical of
the galactic emission for the field.

The low-contrast structure in the 10.8--11.9 $\mu$m image includes
emission near the southern and northeastern
parts of the radio shell. These enhancements are not coincident with the OH
maser locations, nor are they coincident with the 12--18 $\mu$m image. 
There is no emission
from the bright radio bar in the northwest. Indeed, it is just possible to see
an outline of much of the remnant as a `shadow' in the 10.8--11.9 $\mu$m image. 
The `shadow' includes the northwestern radio bar and runs all the way from the 
radio bar along the inside edge of the remnant. The `shadow' does not overlap 
with 3C~391:BML;
instead, it is interior to the shock as traced by the radio emission and the
infrared emission lines. The origin of the `shadow' is unclear. It could be
due to extinction, from unshocked gas. 
We suspect that the shadow actually represents a {\it deficit}
of grains in the post-shock material. It is also possible that some faint
PAH emission occurs outside the radio shell.

The 10.8--11.9 $\mu$m image (Fig.~\ref{figlw8}) may help resolve two other issues for 3C~391. 
First, there is an extension from the remnant in the 12--18 $\mu$m image (Fig.~\ref{figlw3},
described in \S 2.1),
stretching to the south and seemingly doubling the infrared remnant size relative
to the radio size. The 10.8--11.9 $\mu$m image is too small to show the entire extension,
but there is emission in the lower-right (southernmost) part of the 10.8--11.9 $\mu$m image,
which overlaps with the extension. Thus, it appears that the extension is
bright in the aromatic features, whereas the remnant emission is faint
in 10.8--11.9 $\mu$m and has no aromatic hydrocarbons.
Second, the 10.8--11.9 $\mu$m image reveals a bright, diffuse peak just north of the eastern
OH maser. This peak may explain why the 12--18 $\mu$m image extends beyond the boundary
of the radio shell at this location; specifically, the 12--18 $\mu$m image contains some
contribution from dust continuum that is correlated with the aromatic features
traced by the 10.8--11.9 $\mu$m image. Both the southwestern extension of the remnant
and the northeastern peak are outside of the radio shell, and they are
bright in both the 12--18 $\mu$m and 10.8--11.9 $\mu$m filters, suggesting they are due 
to material with very different physical conditions from the post-shock material.


The efficient destruction of small grains is actually counter to what we
had expected. Small grains are actually predicted
to be {\it produced} in 100 km~s$^{-1}$ shocks, being shattered fragments of larger
grains \citep{jones96}. 
The far-infrared spectra of molecular supernova remnants revealed dust
continuum that was associated with the remnants; the shocked dust had a
warmer far-infrared color temperature suggesting that the grains are probably 
smaller than the pre-shock grains. The smallest grains, which would emit
in the mid-infrared apparently do not continue the size distribution of
grain fragments. The smaller grains can be destroyed by sputtering 
in the hot gas behind the shock, and especially fast collisions can vaporize grains
rather than shatter them. 


\section{Conclusions}


Our new near- and mid-infrared observations have separated the 
molecular and ionic shocks in 3C~391. The ionic shocks, traced
by [\ion{Fe}{2}], [\ion{Ne}{2}], and [\ion{Ne}{3}], follow the
nonthermal radio shell and are due to shocks propagating into
moderate-density gas. The ionic shock morphology is filamentary.
There is no continuum or aromatic spectral features in the 
mid-infrared spectrum of the ionic shocks, despite the
presence of bright aromatic features and continuum in the
mid-infrared spectrum of regions just outside the shock.
Thus the small dust grains and aromatic hydrocarbons are
efficiently destroyed in the ionic shocks, contrary to our
expectation that small grains would be abundantly produced 
as fragments of larger grains.

Comparing our new observations of 3C~391 to the well-studied
remnant IC~443 \citep{rho443}, we see they are generally similar, in that parts
of the remnant shells are dominated by ionic shocks (northwest 
in 3C~391, northeast in IC~443) and parts are dominated by
molecular shocks (southeast in 3C~391, south in IC~443). For
3C~391, the ionic shocks arise from the location where the
supernova remnant is tangent to the surface of a giant molecular cloud,
while the molecular shocks are outside the giant molecular cloud.
For IC~443, it has long been assumed that the pre-shock molecular
cloud is to the south, where the molecular ridge is located.
Our observations show that supernova shocks into a molecular cloud
can be ionic shocks through material with density of
order $10^2$ to $10^3$ cm$^{-3}$. Such preshock material 
is traced by CO($1\rightarrow 0$) emission and defines `molecular clouds' 
as traditionally observed. 
In another molecular-interacting remnant, HB~21, the molecular shocks were
also found to be at locations other than the interface between the
supernova remnant and molecular cloud \citep{koohb21}.
The shocks for which H$_2$ emission
dominates the cooling are those into denser material, which has
a small filling factor and may or may not
be present at the location and time when
the blast wave passes through the molecular cloud. In this sense,
molecular-interacting supernova remnants provide a unique perspective
into the structure of molecular clouds.

The region 3C~391:BML, where tracers of molecular shocks including
an OH 1720 MHz maser, H$_2$O and OH far-infrared emission, and
wide CO and CS mm-wave lines, resolves into a cluster of
partially resolved clumps in the near-infrared H$_2$ image. 
The morphology of the molecular shocks is completely different
from that of the ionic shocks. The ISOCAM images show that the
molecular emission is confined to the 3C~391:BML region, with the
ionic shocks following the radio shell and perhaps even dimming
in the region of bright molecular emission.
One of the molecular clumps corresponds precisely with the location
of the OH maser. The clumps most likely are dense cores from the
giant molecular cloud within which the progenitor of 3C~391 formed.
They are now revealed in H$_2$ emission because of the shocks
that are being driven through them by the supernova explosion some
$10^4$ yr ago. These dense cores have their outer layers stripped
and have somewhat altered chemistry in the shocked regions,
potentially impacting their future star formation potential and
the type of stars they may form.

\acknowledgements 
This work is based on observations obtained at the Hale Telescope, 
Palomar Observatory, 
as part of a continuing collaboration between the California Institute
of Technology, NASA/JPL, and Cornell University.
This work is also based on observations with ISO, an ESA project with instruments funded by ESA Member States
 (especially the PI countries: France, Germany, the Netherlands and the
United Kingdom) with the participation of ISAS and NASA.
This publication makes use of data products from the Two Micron All Sky Survey, which is a joint project of the University of Massachusetts and the Infrared Processing
and Analysis Center/California Institute of Technology, funded by the National Aeronautics and Space Administration and the National Science Foundation.

\end{document}